
\input harvmac.tex
\def\shalf{{\scriptscriptstyle{1\over2}}}
\def\quart{{1\over4}}
\def\sign{{\rm sign}}

\def\apm{\alpha^{\prime}}

\def\exp{{\rm exp}}

\def\sinh{{\rm sinh}}
\def\cosh{{\rm cosh}}

\def\tint{\int_{-T/2}^{T/2}}
\def\tisint{\int\prod_{i=1}^n dt_i}
\def\vX{\vec X}
\def\nsum{\sum_{n=0}^\infty}
\def\dtaus{d\tau_1\dots d\tau_n\,}
\def\epiqxtau#1{e^{i \vec q_{#1} \cdot \vX (\tau_{#1})}}

\def\epiqx0#1{e^{i \vec q_{#1} \cdot \vX_0}}
\def\eiqxt#1{e^{i q_{#1} X(t_{#1})}}
\def\eqxtau#1{e^{i q_{#1} X(\tau_{#1})}}
\def\aab{{\alpha \over \alpha^2 + \beta^2}}
\def\bab{{\beta \over \alpha^2 + \beta^2}}
\def\pibab{{2\pi\beta \over \alpha^2 + \beta^2}}
\def\In{{\prod_{i<j}(t_i-t_j)^2 (s_i-s_j)^2\over\prod_{i,j}(t_i-s_j)^2}}
\def\lb{\left\langle}
\def\rb{\right\rangle}

\lref\azbelone{M. Ya. Azbel', Zh.Eksp.Teor.Fiz. {\bf 46}, (1964) 929 [Sov.Phys.
JETP {\bf 19}, (1964) 634].}
\lref\hof {D. R. Hofstadter, Phys. Rev. {\bf B14} (1976) 2239.}
\lref\wannier{G. H. Wannier, Phys. Status Solidi  {\bf B88} (1978) 757.}
\lref\cgcdef{C. G. Callan and D. Freed,
``Phase Diagram of the Dissipative Hofstadter Model'',
Princeton Preprint PUPT-1291, June 1991.}
\lref\dissqm{A. O. Caldeira and A. J. Leggett, Physica  {\bf 121A}(1983) 587;
Phys. Rev. Lett. {\bf 46} (1981) 211; Ann. of Phys. {\bf 149} (1983) 374.}
\lref\osdqm{C. G. Callan, L. Thorlacius, Nucl. Phys. {\bf B329} (1990) 117.}
\lref\cardy{J. L. Cardy, J. Phys. {\bf A14} (1981) 1407.}
\lref\kjaer{K. Kjaer and H. Hillhorst,
Journal of Statistical Physics {\bf 28} (1982) 621.}
\lref\clny{C. G. Callan, C. Lovelace, C. R. Nappi, and S. A. Yost,
Nucl. Phys. {\bf B293} (1987) 83; Nucl. Phys. {\bf B308} (1988) 221.}
\lref\CTrivi{C. G. Callan, L. Thorlacius, Nucl.  Phys. {\bf B319} (1989) 133.}
\lref\fisher{M. P. A. Fisher and W. Zwerger, Phys. Rev. {\bf B32} (1985) 6190.}
\lref\klebanov{I. Klebanov and L. Susskind, Phys. Lett. {\bf B200} (1988) 446.}
\lref\ghm{F. Guinea, V. Hakim and A. Muramatsu,
Phys. Rev. Lett.  {\bf 54} (1985) 263.}
\lref\afflud{I. Affleck and A. Ludwig, Phys. Rev. Lett {\bf 67} (1991) 161.}
\lref\tseyt{E. Fradkin and A. Tseytlin, Phys. Lett. {\bf 163B} (1985) 123.}
\lref\abou{A. Abouelsaood, C. G. Callan, C. R. Nappi and S. A. Yost,
Nucl. Phys. {\bf B280} (1987) 599.}
\lref\defjah{D.~Freed and J.~A.~Harvey, Phys. Rev. {\bf B41} (1990) 11328.}


\Title{\vbox{\baselineskip12pt
\hbox{PUPT-1292}\hbox{CTP~\#2073}\hbox{hepth@xxx/9202085}}}
{Critical Theories of the Dissipative Hofstadter Model}
\centerline{Curtis G. Callan{$^\dagger$},
		\quad Andrew G. Felce\footnote{$^\dagger$}
{callan@puhep1.princeton.edu,~~felce@puhep1.princeton.edu}}
\centerline{\it Department of Physics, Princeton University}
\centerline{\it Princeton, NJ 08544}
\vskip .2in
\centerline{Denise E. Freed\footnote{$^{\dagger\dagger}$}
{freed@mitlns.mit.edu}}
\centerline{\it Center for Theoretical Physics, MIT}
\centerline{\it Cambridge, MA 02139}
\vskip .3in
\centerline{\bf Abstract}
\smallskip
It has recently been shown that the dissipative Hofstadter model
(dissipative quantum mechanics of an electron subject to uniform
magnetic field and periodic potential in two dimensions) exhibits
critical behavior on a network of lines in the dissipation/magnetic
field plane. Apart from their obvious condensed matter interest, the
corresponding critical theories represent non-trivial solutions of open string
field theory, and a detailed account of their properties would be interesting
from several points of view. A subject of particular interest is the dependence
of physical quantities on the magnetic field since it, much like
$\theta_{\rm QCD}$, serves only to give relative phases to different sectors of
the partition sum. In this paper we report the results of an initial
investigation of the free energy, $N$-point functions and boundary
state of this type of
critical theory. Although our primary goal is the study of the magnetic
field dependence of these quantities, we will present some new results
which bear on the zero magnetic field case as well.

\Date{2/92}

\newsec{Introduction}

The Hofstadter problem concerns the quantum mechanics of an electron
moving in two dimensions subject to a magnetic field and a periodic
potential.  The energy bands of this model show a remarkable fractal
structure \refs{\azbelone, \hof, \wannier} as a function of the number
of flux quanta per lattice unit cell.  In a previous paper \cgcdef, the
Caldeira-Leggett model \dissqm\ of dissipative quantum mechanics (DQM)
was used to study how this discontinuous behavior is smoothed out by the
unavoidable elements of randomness in real physical systems.  A
complicated phase diagram was discovered which showed precisely how this
``smoothing out'' works: Above a certain critical dissipation, the particle
is localized, but, as the dissipation is reduced, there is an increasingly
dense system of phase-transition lines which have a fractal structure
in the zero-dissipation limit.  The topic of this paper is the study of
the properties of the one-dimensional critical theories corresponding to
the phase transitions themselves.  In addition to their relevance to the
Hofstadter problem, these theories represent new solutions of open
string theory in a non-trivial background of tachyons and gauge fields
\osdqm .  The string theory connection strongly suggests that the critical
theory should have an enhanced $SL(2,R)$ invariance in addition to
the usual scale invariance and we will verify that this is so.

We have found a simple regulator which reduces the calculation of most
quantities of interest, to any order of perturbation in the potential, to
the purely algebraic exercise of extracting the residues of poles in a
rational function associated with each perturbation theory diagram.
Using this regulator, for some special points in the phase diagram
we explicitly demonstrate the absence of logarithmic divergences, and
show that the only renormalization needed is a rescaling of the
potential strength and the subtraction of an infinite constant from the free
energy. We give evidence for the criticality of the circular arcs in the phase
diagram which join these special points to one another, although we
cannot give a proof to all orders in this case. In addition we have
done some explicit calculations of the magnetic field dependence of free
energies and $N$-point functions to various orders in the potential strength.
We find that many of the connected higher $N$-point functions are zero,
up to contact terms. Although this suggests that the critical theories are
``almost'' free and therefore ought to be soluble, we have not been able to
exploit this hint to obtain exact solutions and must for the moment content
ourselves with the rather clumsy perturbative approach presented here.

In the interests of making this paper relatively self-contained, we devote
the first two sections to a brief review of background material that
has, by and large, appeared elsewhere.  In Section 2 we
review dissipative quantum mechanics and its relationship with open string
theory.  In Section 3 we specialize to the Hofstadter model.  We show
that it is equivalent to a Coulomb gas, demonstrate that it has phase
transitions and show that the critical theories have $SL(2,R)$ invariance.
We then move on to the considerations which are new to this paper.
In Section 4 we give a fairly complete discussion of the unique critical
theory at zero magnetic field, and show that it is a free theory with
certain nontrivial contact interactions. In Section 5 we calculate the
free energies and $N$-point functions of critical theories at non-zero
magnetic field in the zero-charge sector of the Coulomb gas.
An interesting feature of our results is that at the previously-mentioned
special points in the phase diagram, most of the $N$-point
functions reduce to contact terms.  There are, however, some few that
do not and we give their $SL(2,R)$-invariant form.  Section 6 is devoted
to the non-zero charge sectors of the Coulomb gas.  They are crucial to
the construction of the open string boundary state, and we find that
no new renormalizations are needed, beyond those needed to deal with the
zero-charge sector.  Our conclusions are in Section 7.  In one appendix
we present details of the proof that all the higher $N$-point functions
reduce to contact terms in the absence of a magnetic field.  In
another, we show that the $N$-point functions satisfy the very
non-trivial string theory reparametrization invariance Ward identities.

\newsec{Background: Dissipative Hofstadter Model and Open String Theory}

We begin with a brief outline of dissipative quantum mechanics and its
connection with open string theory.  For details the reader is referred to
\dissqm\ and \osdqm.  A macroscopic object is typically subject to
dissipative forces caused by its interaction with its environment.
Classically, these forces can be described by including the phenomenological
term $-\eta \dot X^i$ in the equation of motion for the particle, where
$\eta$ is the coefficient of friction and $X^i$ the particle's
coordinate.  In order to describe dissipation quantum mechanically, one can
model the environment by a bath of an infinite number of harmonic
oscillators coupled linearly to the ${\vX}$.  The coupling constants,
$C_\alpha$, and the distribution of the frequencies, $\omega_\alpha$,
of the oscillators can be
chosen so that when the oscillator coordinates are eliminated via their
equations of motion, the resulting equations of motion for the $X^i$
contain the required friction term.  The functional condition on the
parameters is
\eqn\parcon{J(\omega)=\pi\sum_\alpha {C_\alpha^2 \over 2\omega_\alpha}\;
           \delta (\omega {-}\omega_\alpha ) = \eta\,\omega~,}
which represents Ohmic dissipation in the system.
This is the Caldeira-Leggett model \dissqm.

Since the dependence on the oscillator coordinates in
the lagrangian is quadratic, it is also possible to integrate them out
from the quantum mechanical path integral.  This results in a
quantum effective action for the $X^i$ variables, which includes
a non-local piece containing the effect of dissipation.  For a
particle coupled to scalar ($V$) and vector ($\vec A$) potentials the full
action reads
\eqn\action{S[\vX] = \int dt\bigl\{{\half}M\dot{\vX}^2 + V(\vX)
                           +i\, A_i (\vX)\,\dot X^i \,
                   + {\eta\over 4\pi}\int\limits_{-\infty}^\infty
   dt'\; {\bigl(\vX(t){-}\vX(t')\bigr)^2 \over (t{-}t')^2}\bigr\}\, .}
Remarkably, the only dependence on the oscillator parameters that remains in
the action is through the $\eta$-term.
Because this term is non-local, the path integral
is effectively that of a one-dimensional statistical system
with long-range interactions. Such systems, unlike one-dimensional
{\it local} systems, have phase transitions (the classic example being
the Ising chain with $1/r^2$ interactions \foot{The Dyson chain with
$1/r^n$ interactions for $n\le 2$ is known to have phase transitions.
See references \refs{\cardy, \kjaer} and references therein.}).
In the DQM context, the phase transitions are between different
regimes of long-time behavior of Green's functions (typically
between localized and delocalized behavior).

At these critical points, the 1-D field theories describing dissipative
quantum mechanics correspond to solutions of open string theory
\osdqm.  In the presence of open string background fields,
interactions between a string and the background take place at the
boundary of the string and their effects can be represented by a
boundary state $|B\rangle$.  In \clny\ it is shown that this boundary
state is given by
\eqn\bs{|B\rangle=\exp\left\{\sum_{m=1}^\infty{1\over m}\alpha_{-m}\cdot
          \tilde\alpha_{-m}\right\}
       \int\left[D\vX(s)\right]'
		\exp(-S_R -S_{KE} -S_{I} -S_{LS})|0\rangle,}
where
\eqn\SRdef{S_{R}=\int_0^T ds\,\half M\dot{\vX}^2(s)~;}
\eqn\SKEdef{S_{KE}={1\over8\pi^2\apm}\int_0^T ds\int_{-\infty}^\infty ds'
      {\bigl(\vX(s)-\vX(s')\bigr)^2\over(s-s')^2}~;}
\eqn\SIdef{S_I=i\int_0^T ds\,A_\mu(\vX)\dot X^\mu
        +\int_0^T ds\, {\cal T}(\vX)~;}
and
\eqn\SLSdef{S_{LS}=\sqrt{2\over\alpha'}\int ds~\alpha(s)\cdot\vX(s)
      \qquad {\rm with}\quad
      \alpha^\mu(s)=\sum_{m=1}^\infty i(\tilde\alpha_{-m}^\mu e^{-ims}+
           \alpha_{-m}^\mu e^{ims})~.}
In these expressions, $T$ is the parameter length of the boundary
(when appropriate, we must regard $\vX(s)$ as periodic in $s$ with
period $T$), $\apm$ is the string constant and $A_\mu(\vX)$ and
${\cal T}(\vX)$ are the gauge fields and tachyon fields, respectively.
The creation operators of the left- and right-moving modes of the closed
string, $\tilde\alpha_{-m}$ and $\alpha_{-m}$, act on the closed string
vacuum $|0\rangle$ to create some state in the closed string Hilbert
space.  The notation $\bigl[D\vX(s)\bigr]'$
means that the zero-mode, $\vX_0$ is not integrated out. The commuting
objects $\tilde\alpha_{-m}$, $\alpha_{-m}$ and $\vX_0$ together make up a
set of coordinates which specify where the boundary lies in the target space
and the boundary state is just a functional of these coordinates.
As an example of the utility of this construct, we note that the projection
of $|B\rangle$ onto the graviton state is essentially the energy-momentum
tensor of the open string object under study.  This gives us a
string-theoretic way to define such important notions as gravitational
and inertial mass.

This path integral is the generating functional for a renormalizable
``one-dimensional'' field theory described by the underlying action
$S_{KE}+S_I$:
($S_{LS}$ is the linear source term in the generating function and
the kinetic term $S_R$ functions as a regulator for divergences).
Leaving aside the linear source term, it is clear that the DQM action and
the action defining the string theory boundary state are the same if we
relate the coefficient of friction to the string tension by
$\eta=1/(2\pi\alpha')$. The full string theory prescription for $|B\rangle$
requires that we take the cut-off, $M$, to zero. In order for this limit to
be meaningful, the field theory must lie at a renormalization group fixed
point, which is to say that the gauge and tachyon fields must satisfy some
``vanishing beta function'' equations of motion for open string background
fields.  This means that the associated DQM must lie at a phase
transition.  The upshot is that, modulo technical  details, any solution
of the open string equations of motion is  equivalent to a particular
critical DQM: The background gauge and tachyon fields of the string theory
become the vector and scalar potentials to which the DQM electron is
subject.

In string theory we require worldsheet reparametrization invariance, and
this includes reparametrizations of the boundary. The condition that
the boundary state be reparametrization invariant can be written
\eqn\vira{(L_n - \tilde{L}_{-n})|B\rangle = 0 ~.}
where the $L$ operators are the closed string Virasoro generators (they act
in a known way on the coordinates $\alpha_{-m}$, $\tilde\alpha_{-m}$ and
$\vX_0$ on which the boundary path integral depends). This symmetry
generates a set of Ward identities which turns out to be very
useful both in string theory and in DQM (details can be found in \CTrivi).
In fact, the one-dimensional field theory has what amounts to {\it broken}
reparametrization invariance, because of the non-local
dissipation term.  There is nonetheless a remaining manifest $SL(2,R)$
symmetry ($SU(1,1)$ on the circle) which tightly constrains the allowed
form of the correlation functions. These extra symmetries would not
have been expected at the one-dimensional critical points without the
string theory connection and we will look for them in our explicit
calculations.

\newsec{General Properties of the Dissipative Hofstadter Model}

\subsec{Equivalence to a Coulomb Gas}

To specialize to the dissipative Hofstadter model, we consider a particle
moving in a periodic potential in two dimensions (with $\vX=(X,Y)$) and
subject to a uniform magnetic field $B$.  The Euclidean action for this
problem is the sum of a quadratic piece and a more complicated potential
term:
\eqn\Sdef{S = S_q + S_V}
where
\eqn\Sqdef{{1\over\hbar} S_q ={1\over\hbar}\tint dt\bigl\{
		{M\over2}\dot{\vX}^2+
		{ieB\over2c}(\dot X Y - \dot Y X)+
		{\eta\over4\pi}\int\limits_{-\infty}^\infty
		dt'\, {(\vX(t) - \vX(t'))^2 \over (t-t')^2}\bigr\}}
and
\eqn\SVdef{S_V = \tint dt~V(X,Y)~.}
For the periodic potential we take
\eqn\Vdef{V(X,Y) = -V_0\cos({2\pi X(t)\over a})
                   -V_0\cos({2\pi Y(t)\over a})~.}
Nothing dramatically new happens if we take the strength and
period of the potential to be different in the X- and Y- directions.
It is convenient to define the dimensionless parameters
\eqn\defalpha{2\pi\alpha = {\eta a^2 \over \hbar},\qquad
		2\pi\beta = {eB\over \hbar c} a^2~,}
to rescale $X$ and $Y$ by $a/2\pi$, and to rescale $V_0$ by $\hbar$.
Until we come to consider the infrared regulation of the theory in
Section 3.4, we take $T$ to be infinite, which means that we are at zero
temperature, and that the particle lives on a line.
Then the action $S_q$ can be written as
\eqn\Sqtwo{{1\over\hbar} S_q = \half\int {d\omega \over 2\pi}
                     \left\{\left({\alpha\over 2\pi} |\omega| +
                    {Ma^2\over \hbar} \omega^2 \right) \delta_{\mu\nu} +
                   {\beta \over2\pi}\epsilon_{\mu\nu}\omega\right\}
                    \tilde X^*_\mu(\omega) \tilde X_\nu(\omega)~.}

Because the ordinary kinetic term, $\half M\dot{\vX}^2$, is a dimension-two
operator, it is irrelevant and acts only as a regulator as far as the
large-time behavior is concerned. Since we are studying critical behavior,
it will be legitimate to set $M=0$ and use some other, more convenient,
regulator where needed. The Fourier-transformed propagator defined by
\Sqtwo\ (with $M=0$) is
\eqn\Gwdef{G^{\mu\nu}(\omega) = 2\pi\aab {1\over |\omega|} \delta^{\mu\nu}
                     - 2\pi \bab {1\over \omega} \epsilon^{\mu\nu}.}
In the time domain, this becomes
\eqn\Gtdef{G^{\mu\nu}(t_i-t_j) = -\aab \ln(t_i-t_j)^2 \delta^{\mu\nu}
                      - {i\over 2} \pibab \sign(t_i-t_j) \epsilon^{\mu\nu}.}
For future reference, we note that in one dimension and for vanishing magnetic
field this propagator reduces to
\eqn\GtnoBdef{G(t_1-t_2) = - {1\over\alpha} \ln(t_i - t_j)^2.}

Except for the cosine potential, the action \Sdef\ is quadratic, so we will
treat the potential as a perturbation. We proceed by expanding
\eqn\Vexp{\exp\left(V_0\int\cos X(t)dt\right) =
   \nsum \int\dtaus \left({V_0\over 2}\right)^n
   {1\over n!} \sum_{q_j = \pm1}\prod_{j=1}^n \eqxtau j .}
Then the partition function is given by
\eqn\Zdef{\eqalign{Z &= \exp\left(-{1\over \hbar} S\right) \cr
  &= \int D\vX(t) \nsum \int \dtaus
    \left({V_0\over 2}\right)^n {1\over n!}
    \!\sum_{\vec q_j ={(\pm 1,0)\atop(0, \pm1)}}
   \prod_{j=1}^n \epiqxtau j e^{-S_q/\hbar}\cr
  &=  \nsum \int \dtaus \left({V_0\over 2}\right)^n {1\over n!}
		\sum_{\vec q_j}
     \lb\prod_{j=1}^n \epiqxtau j \rb_0,}}
where the functional integral is over periodic paths, and the
correlation functions of the operators $e^{i \vec q \cdot \vX(t)}$
are to be computed with the propagator \Gtdef . A subtlety to be borne
in mind is that for the dissipative quantum mechanics system, we must
integrate over the zero mode in equation \Zdef, which imposes the charge
conservation requirement $\sum \vec q_i = 0$.  For the string theory
boundary state path-integral, we omit the integration over the zero
mode, and the $q_i$'s are unconstrained.

If we restrict to one dimension and drop the magnetic field, the $O({V_0}^n)$
term in \Zdef\ is a sum over $q_j = \pm 1$ of
\eqn\Zndef{Z_n = {1\over n!}\left({V_0\over2}\right)^n \tisint \lb
                 \eiqxt 1 \eiqxt 2 \dots \eiqxt n \rb_0}
Using \GtnoBdef, this evaluates to
\eqn\Znln{Z_n = {1\over n!} \left({V_0\over 2}
     e^{-\shalf\lb X^2(0)\rb}\right)^n \tisint~
     \exp\bigl({1\over\alpha}\sum_{i<j}q_i q_j \ln(t_i-t_j)^2\bigr)~.}
The expression in the exponent is the free energy for $n$ particles with
charges $q_j=\pm 1$, interacting logarithmically and restricted to a
line.  For the dissipative quantum mechanics
system, the condition that $\sum q_j = 0$ just requires the gas to be
neutral.  For the boundary state path integral, the gas can have any
charge. The full partition function describes a one-dimensional Coulomb
gas of particles with fugacity proportional to $V_0$.

Similarly, in the case of most interest to us (two dimensions and non-zero
field), the $O({V_0}^n)$ term for $Z$ is a sum over $\vec q_j = (\pm1, 0)$ and
$(0, \pm1)$ of
\eqn\ZnBln{\eqalign{Z_n = &{1\over n!} \left({V_0\over 2}
          e^{-\shalf\lb X^2(0)\rb}\right)^n \tisint\cr
         & \exp\sum_{i<j}\left\{\aab \vec q_i \cdot \vec q_j
           \ln(t_i-t_j)^2
           +{i\over 2} \pibab \epsilon^{\mu\nu} q_i^\mu q_j^\nu
            \sign(t_i - t_j)\right\}.}}
This has an interpretation as a generalized ``Coulomb'' gas.  Now there
are two species of particles, one corresponding to the X-component of
$\vec q$ and one to the Y-component of $\vec q$.  Charges of
the same species still interact logarithmically (through the $\delta_{\mu\nu}$
piece of \Gtdef). Charges of differing species only interact through a sign
function (the $\epsilon_{\mu\nu}$ piece of \Gtdef) and the wave-function
picks up a phase factor when they are interchanged.  One of our main points is
that this simple generalization of the Coulomb gas has a very rich phase
structure.

This Coulomb gas sum has an additional interpretation.  When there is
no dissipation, it was shown in ref. \defjah\ that the partition function
\ZnBln\ describes an electron in a Landau orbit centered on a dual lattice
site, $m (a/\beta)\hat x+ n (a/\beta)\hat y$.  Whenever the potential acts
(via an insertion of a $(V_0/2)e^{i\vec q_j \cdot \vec X(t_j)}$), the center
of the Landau orbit hops by $\vec q_j$ units in the reciprocal lattice and the
action picks up the Aharonov-Bohm phase due to this hop.  Once we turn on
the dissipation, according to equation \ZnBln, there are two major changes.
The first is that the dual lattice becomes
\eqn\reclat{m {a\over\alpha^2 + \beta^2}\hat x
             + n {a\over\alpha^2 + \beta^2}\hat y~,}
so that now, when the center of the particle's orbit hops through a square
in the reciprocal lattice, it picks up a phase of
$\pm 2\pi\beta/(\alpha^2 + \beta^2)$.  Additionally, there is now a
logarithmic interaction between the particle's hops at time $t_i$ and time
$t_j$.

\subsec{Phase Transitions in the Dissipative Hofstadter Model}

In standard quantum mechanics, a particle in a periodic potential is
delocalized by coherent quantum tunneling effects. In the presence of
strong enough dissipation one expects coherence between tunneling events
to be lost and the particle to become localized. The signature of this
localization-delocalization phase transition can be looked for in the
asymptotic behavior of the two-point function, $\lb X(t_1)X(t_2)\rb$: Let
the long-time behavior of the two-point function be $(t_1-t_2)^\gamma +$
const.  Localization corresponds to $\gamma <0$, delocalization to
$\gamma > 0$. When $\gamma=0$, so that
$\lb X(t_1)X(t_2)\rb\sim\ln(t_1-t_2)$, the system is at a critical point.

Fisher and Zwerger \fisher\ have given a renormalization group
argument which shows that DQM with a periodic potential and zero magnetic
field is critical at $\alpha=1$.
Their argument is valid for small values of the potential strength, $V_0$,
since they treat the potential perturbatively, but sum over all loops.
Following this calculation closely,  we can extend it to include a constant
magnetic field.  We will present here the calculation to first order in
$V_0$ in order to identify a candidate for a critical circle on the
$\alpha$-$\beta$ plane.  We can show that the results given here
remain true at order $V_0^2$, and we expect that they will hold for all
orders in $V_0$.

In the action defined in \Sdef\ and the following equations, we
set the mass term to zero and regulate with a high frequency cut-off,
$\Lambda$, instead.  The action is then given by
\eqn\SLambda{{1\over\hbar}S=\half\int_{-\Lambda}^\Lambda
            {d\omega\over2\pi}
           \tilde X_\rho^\dagger(\omega)S^{\rho\lambda}(\omega)
           \tilde X_\lambda(\omega)+\Lambda V_0(\Lambda)\int
		d\tau\left[\cos X(\tau)+\cos Y(\tau)\right],}
with
\eqn\SmunuLambda{S^{\mu\nu}(\omega)
               ={\alpha\over2\pi}|\omega|\delta^{\mu\nu}
               +{\beta\over2\pi}\omega\epsilon^{\mu\nu}}
and $\Lambda V_0(\Lambda)$ replacing the bare coupling $V_0$.
At this point, we want to perform the functional integral over all the
fast modes, $\tilde X(\omega)$, with $\Lambda>\omega>\mu$, for some
$\mu<<\Lambda$.  To do so, we divide the field into fast and slow modes,
\eqn\XsXf{\eqalign{\vX(\tau)=&~\vX_s(\tau)+\vX_f(\tau)~,\qquad
            {\rm where}\cr
          \vec{\tilde X}(\omega)\approx&~\vX_s(\omega) \qquad
            {\rm for} \quad |\omega|\le\mu \qquad {\rm and}\cr
          \vec{\tilde X}(\omega)\approx&~\vX_f(\omega) \qquad
            {\rm for} \quad \mu\le|\omega|\le\Lambda~.}}
We would like to calculate $\tilde S$, where $\tilde S$ is defined by
the relation
\eqn\Stildedef{\int D\vX(\tau) e^{-{1\over\hbar}S}
               = \int D\vX_s(\tau) e^{-{1\over\hbar}\tilde S}~.}
The coefficients of $|\omega||\vec{\tilde X}(\omega)|^2$,~
$\omega\epsilon^{\sigma\nu}\tilde X_\sigma^*(\omega)\tilde X_\nu(\omega)$,
and $\cos X_s(\tau) + \cos Y_s(\tau)$ in $\tilde S$ determine the flows
of $\alpha$, $\beta$ and $V_0$, respectively.  To calculate $\tilde S$,
we treat the potential term perturbatively, just as we did to obtain
equation \Zdef.  It is not too hard to show that, to first order in $V_0$,
\eqn\tildeScalc{\tilde S=\half\int_{-\mu}^{\mu}
            {d\omega\over2\pi}
           \tilde X_\rho^\dagger(\omega)S^{\rho\lambda}(\omega)
           \tilde X_\lambda(\omega)
           +\lb S_V\rb_f~.}
where
$S_V=\Lambda V_0(\Lambda)\int d\tau\left[\cos X(\tau) +\cos Y(\tau)\right]$,
and the gaussian average over the fast modes is calculated with the
following two-point function:
\eqn\Gfast{G^{\rho\lambda}(\tau)
           =\lb X_f^\rho(\tau) X_f^\lambda(0)\rb
           = \int_{-\Lambda}^\Lambda d\omega
		\left(S^{-1}\right)^{\rho\lambda}(\omega)
                         e^{i\omega\tau}W(\omega/\mu)~.}
$W(x)$ must be $\approx 0$ for $x<<1$ and $\approx 1$
for $x >>1$.  To the order in $V_0$ we are calculating here, it is
sufficient to take $W(x)$ to be a step function.  (If we wish to continue the
calculation to higher orders in $V$, then $W(x)$ must be smooth enough to
avoid the generation of spurious long-range behavior in $G(\tau)$.)
With our choice for $W(x)$, we can calculate the diagonal part of
$G^{\rho\lambda}(0)$ with the result
\eqn\GLambda{G^{\rho\lambda}(0)
             ={2\alpha\over\alpha^2+\beta^2}\ln(\Lambda/\mu)
                        \qquad {\rm for}\quad \rho=\lambda~.}
$\lb S_V\rb_f$ can be written solely as a function of $G^{XX}(0)=G(0)$ as
follows:
\eqn\SVcalc{\lb S_V\rb_f=\Lambda V_0(\Lambda)e^{-\shalf G(0)}
         \int d\tau\,\left[\cos X_s(\tau)+\cos Y_s(\tau)\right]~.}
Finally, we rescale $\tau$ by $\Lambda/\mu$ to restore $\tilde S$ to its
original form and find that the coefficient of the potential term becomes
\eqn\Vflow{\eqalign{V_0(\mu)=&
             V_0(\Lambda)(\Lambda/\mu)e^{-\shalf G(0)}\cr
            =& V_0(\Lambda)(\mu/\Lambda)^{(\aab-1)},}}
while the coefficients of the friction term and magnetic field term do
not change.  We conclude that, as we take $\mu/\Lambda$ to $0$, the
potential term flows to zero if $\alpha/(\alpha^2+\beta^2)>1$, remains
fixed when $\alpha/(\alpha^2+\beta^2)=1$ and grows when
$\alpha/(\alpha^2+\beta^2)<1$.  Also, to this order in $V_0$, we
have just shown that $\alpha$ and $\beta$ do not flow.  Therefore, we
have demonstrated the existence of a critical circle in the
$\alpha$-$\beta$ plane, to first order in $V_0$.
We note that we obtain a critical circle for {\it any}
initial value of $V_0$, as long as $V_0$ is small enough to
justify the perturbative expansion. Inside this circle, $V_0$ is
irrelevant, so the particle should be delocalized, and, outside the
circle, $V_0$ is relevant.

We expect this behavior to continue at higher orders in $V_0$ because the
only relevant terms that are generated in $\tilde S$ are of the form
$\cos X_s(\tau) + \cos Y_s(\tau)$.  In particular, the non-local kinetic
term and the magnetic term are not generated, so the friction per unit cell
and the magnetic flux cannot flow.  Thus, we expect the circle
$\alpha^2 + \beta^2$ to be critical to all orders in $V_0$.

\subsec{Symmetries of the Critical Theory}

On the critical circle, where $\aab = $1, the unregulated theory has
several nice properties. First, the dissipative quantum mechanics
system displays $SL(2,R)$ invariance, not just scale invariance.
As mentioned in Section 2, we expected this larger symmetry group at the
phase transition because of the connection with open string theory. This
invariance means that, under the transformation
\eqn\mobius{t \rightarrow \tilde t = {at + b \over ct + d}
              \qquad ad -bc = 1 \qquad a,b,c,d, \in R~,}
the form of the partition function remains unchanged while
$\dot X(t)$ transforms as a dimension one operator and $e^{ikX(t)}$
transforms as a dimension $k^2$ operator.

To show this, we first consider the $O({V_0}^{2n})$ term of the partition
function for a neutral gas. For simplicity, we study the zero-field case.
Note, however, that the $SL(2,R)$ invariance remains when the magnetic field is
non-zero, because the magnetic-field-dependent contributions to $Z_n$ and to
the
correlation functions depend only on the ordering of the $t_i$, which remains
unchanged under the $SL(2,R)$ transformation.

In the zero-field case ($\beta=0$) one has a critical point at
$\alpha = 1$. At this point, we can rewrite equation \Znln\ for $Z_{2n}$ as
\eqn\Ztwon{Z_{2n} = {1\over(2n)!}\left({V_0\over 2}
                    e^{-\shalf\lb X^2(0)\rb}\right)^{2n}
                    \int \prod_{i=1}^n {dt_i\over 2\pi} {ds_i\over 2\pi}
                    \In.}
This is the $O(V_0^{2n})$ term of the partition function with $+1$
charges at the $t_i$'s and $-1$ charges at the $s_j$'s.
This expression clearly remains unchanged under the translation
$t_i \rightarrow t_i+a$ and $s_j \rightarrow s_j +a$.
It also remains unchanged when all the variables are rescaled by
the same factor. The third transformation needed to generate $SL(2,R)$
can be taken to be $t \rightarrow -1/t$, and it can easily be
seen that $Z_{2n}$ is invariant under this inversion as well. Thus, the
partition function is invariant under $SL(2,R)$ transformations.

Operator $N$-point functions are only slightly more complicated.
Using the properties of gaussian propagators, one can show that the
connected correlation functions
$\langle\dot X(r_1)\dots\dot X(r_m)\rangle$
are given by the connected part of the partition function with the
insertion of
\eqn\twoptin{\eqalign{(-1)^{m/2}\prod_{j=1}^m\Biggl\{\sum_i
                        \lb\dot X(r_j) X(t_i) \rb_0 -&
                  \lb\dot X(r_j) X(s_i) \rb_0 \Biggr\}\cr
             =& (-1)^{m/2}\prod_{j=1}^m\left\{\sum_i
               \left({2\over r_j -t_i} - {2\over r_j-s_i}\right)\right\}\cr
             =& (-4)^{m/2}\prod_{j=1}^m\left\{\sum_i
                {t_i-s_i \over (r_j-t_i)(r_j-s_i)}\right\}.}}
Because $Z_{2n}$ is invariant under $SL(2,R)$ transformations, we only
need to see how the expression in curly brackets transforms under
$r_j \rightarrow \tilde r_j = (ar_j +b)/(cr_j+d)$ with $ad-bc =1$ (along
with a simultaneous transformation of $t_i$ and $s_i$).
We find
\eqn\transts{{t_i-s_i\over (r_j-t_i)(r_j-s_i)} \rightarrow
        {\tilde t_i-\tilde s_i\over
        (\tilde r_j-\tilde t_i)(\tilde r_j-\tilde s_i)}
        = (c r_j + d)^2 {t_i-s_i\over (r_j-t_i)(r_j-s_i)}~.}
Taking the derivative of $\tilde r$ with respect to $r$, we have
\eqn\drtild{{d\tilde r \over dr} = {1\over (cr+d)^2},}
from which it follows that
\eqn\trancor{\lb\dot X(r_1) \dots \dot X(r_m)\rb =
           \prod_{i=1}^m \left({d\tilde r_i \over d r_i }\right)
           \lb\dot X(\tilde r_1) \dots \dot X(\tilde r_m)\rb~.}
Thus, insertions of $\dot X(r)$ transform as dimension-1 operators.
For correlation functions like
$\lb e^{ik_1X(t_1)}\dots e^{ik_2X(t_m)}\rb$ with $\sum k_j = 0$,
the calculation is similar.

\subsec{The Regulated Theory}

So far, we have only considered the unregulated theory.  However, it has
both infrared and ultraviolet divergences which we must regulate.  The
ultraviolet divergence was originally regulated by the $M\dot X^2$ term,
which acts as a high frequency cutoff by multiplying the Fourier-space
propagator, $\tilde G(\omega)$, by $1/(M|\omega| + \eta)$.  We find it
more convenient to use an $e^{-\delta|\omega|}$ regulator, where
$\delta$ is a dimensionful ultraviolet cutoff.  To take care of the
infrared divergence, we put the particle on a circle of circumference $T$,
which, in Euclidean space, is equivalent to looking at finite temperature.
In this case, the propagator is
\eqn\Gmmdef{\eqalign{G^{\mu\nu}(t_1-t_2)
          =&\aab \sum_{m\neq0}{1\over |m|}e^{im(t_1-t_2)2\pi/T}
          e^{-\epsilon|m|}\cr
        =&-\aab \ln\left(1 + e^{-2\epsilon}-2e^{-\epsilon}
            \cos{2\pi\over T}(t_1-t_2)\right)}}
when $\mu = \nu$ and
\eqn\Gmndef{\eqalign{G^{\mu\nu}(t_1-t_2) =&-\epsilon^{\mu\nu}\bab
           \sum_{m\neq0}{1\over m}e^{im(t_1-t_2)2\pi/T} e^{-\epsilon|m|}\cr
          =&2i \epsilon^{\mu\nu} \bab \arctan\left({\sin[2\pi(t_1-t_2)/T]
                 \over\cos[2\pi(t_1-t_2)/T] - e^\epsilon}\right)}}
for $\mu\ne\nu$.  In these expressions, $\epsilon$ is a dimensionless
cutoff which is the ratio of the ultraviolet cutoff $\delta$ to the
infrared scale $T$.  The expressions for $G^{\mu\nu}$ take on much
simpler forms if we change variables to $z_j = e^{2\pi i t_j/T}$.
With this definition, the exponentiated propagator is
\eqn\eGmn{e^{-q_1^\mu q_2^\nu G^{\mu\nu}(t_1-t_2)}=
	\left[{-e^\epsilon z_1 z_2 \over (z_1 - e^\epsilon z_2)
               (z_1 - e^{-\epsilon} z_2) }\right]^{\zeta_\alpha}
	\left[{z_1 -z_2 e^\epsilon \over z_2 - z_1 e^\epsilon}
              \cdot {z_1 \over z_2} \right]^{\zeta_\beta},}
where the exponents are
\eqn\zetadef{\zeta_\alpha ={-\vec q_1 \cdot \vec q_2 \aab}~,\qquad
	     \zeta_\beta ={-q_1^\mu q_2^\nu \epsilon^{\mu\nu} \bab}~,}
and we have used the identity
\eqn\atanid{\arctan x = {1\over 2i} \ln\left({1+ix\over 1-ix}\right),}
to rewrite the off-diagonal part of $G$.  Using the same variables we
also have
\eqn\xxdotdef{\lb \dot X(t_1) X(t_2) \rb_0
              = \aab\left(-{2\pi i\over T}\right){z_1^2-z_2^2 \over
		(z_1-e^{-\epsilon}z_2)(z_1 - e^\epsilon z_2)}~,}
and we can obtain similar expressions for
$\langle\dot Y(t_1)Y(t_2)\rangle_0$,
$\langle\dot X(t_1)Y(t_2)\rangle_0$, etc.

Now we exploit simplifications which occur on the critical circle,
{\it i.e.} when $\aab = 1$. First, we note that as
$\epsilon\rightarrow 0$ the exponentiated propagator becomes
\eqn\exxcircle{e^{\lb X(t_1) X(t_2)\rb} =
			-{z_1 z_2\over (z_1-z_2)^2}~.}
This is very similar to the expression for $e^{\lb X(t_1) X(t_2)\rb}$
when $t$ lies on a line.  Therefore, we can give a similar argument to the
one illustrating $SL(2,R)$ invariance on a line to show that on the circle,
the theory has $SU(1,1)$ invariance.  This means that under
\eqn\suoodef{z = e^{2\pi i t/T} \rightarrow \tilde z =
               {az+b\over\bar bz+\bar a} \qquad {\rm with}
                \quad  |a|^2 - |b|^2 = 1~,}
the partition function remains invariant, $\dot X(t)$ transforms as a
dimension one operator and so on.  As a result, even though at finite
temperature the system does not exhibit scale invariance, it still
possesses a larger symmetry group, $SU(1,1)$.

In order to consider what happens when $\epsilon\ne0$, we begin by
specializing to the case when $\vec q_i \cdot \vec q_j\aab$
and $q_i^\mu q_j^\nu\epsilon_{\mu\nu} \bab$ are integers for all
$\vec q_i$. This condition occurs on the critical circle
$\aab=1$ at the ``special points'' where $\beta/\alpha\in Z$.
At these points, the partition function and correlation
functions have a very simple form even before we take the regulator to
zero.  This is because $e^{q_i^\mu q_j^\nu \lb X^\mu(t_i) X^\nu(t_j)\rb}$
and $\langle\dot X^\mu(t_i) X^\nu(t_j)\rangle$ are just rational
functions in the $z_i$'s with coefficients of the form $e^{n\epsilon}$.
For example, at the zero-field critical point, where $\alpha=1$ and
$\beta=0$, the regulated version of the integral in \Ztwon\
which gives the $O(V_0^{2n})$ term of the partition function is
\eqn\Qtwonex{Q_{2n}=\oint\prod_{j=1}^n
                \left({dz_j \over 2\pi i }{dw_j\over 2\pi i}\right)
		I_{2n}~,}
where the definition of $I_{2n}$ is
\eqn\Itndef{I_{2n}=(-1)^n{\prod_{i<j}(z_i-e^{\epsilon}z_j)
		(z_i-e^{-\epsilon}z_j)(w_i-e^{\epsilon}w_j)
		(w_i-e^{-\epsilon}w_j)\over
		\prod_{i,j}(z_i-e^{\epsilon}w_j)(z_i-e^{-\epsilon}w_j)}~.}
Because of the ultraviolet regulator, none of the poles lie on the
contour. Furthermore, the denominator of the integrand is factored into
terms that are linear in all the integration variables, $z_i$ and $w_j$.
Consequently, after we perform the contour integral over one variable, the
resulting integrand is once again a rational function in the remaining
variables with a denominator that is completely factored into linear
terms.  This property of the integrand has several important consequences.

The first conclusion is that it is straight-forward, although tedious, to
analytically evaluate any term in the perturbation expansions for the free
energy and correlation functions.  More generally, there is a well-defined
set of rules for integrating any term in the perturbation series. They
tell us how to go from a graph (or integrand)
with $n$ vertices to several graphs with $n-1$ vertices by evaluating the
residues of a rational function.  With these rules, we can program a
computer to calculate correlation functions analytically.  We have done a
few examples with the help of {\it Mathematica}, and some of the results are
given in Section 4.

We can also use these rules to prove that the free energy and correlation
functions have certain properties.  The most important such property is
that there are no logarithmic divergences, which can be seen as follows.
After we have performed all the integrals, we must
obtain a rational function in $e^\epsilon$ for the free energy (or
a rational function of the variables $e^\epsilon$ and $e^{2\pi i r_{j}}$ for
correlation functions of arbitrary numbers of $\dot X(r_j)$ and
$e^{i X(r_j)}$ fields).  This implies that, in the limit as
$\epsilon\rightarrow 0$, we obtain only rational functions in $\epsilon$.
Thus it is obvious that, to all orders in the periodic potential
(at the special points), no logarithmic divergences in $\epsilon$ are
possible. Pole divergences can of course still occur.

This result is actually quite general. The structure of the partition function
integrand will be the same as above for any doubly periodic potential of the
form
\eqn\pot{V=V_X \cos{2\pi X\over a}+V_Y \cos{2\pi Y\over a}
		+{\rm higher~harmonics}}
at the special points $\aab = 1$ and $\bab \in Z$.  It appears that
imposing the discrete symmetry $X\rightarrow X + {na\over 2\pi}$,
$Y\rightarrow Y + {ma\over 2\pi}$ guarantees that there are no
logarithmic divergences at these points in the phase diagram.  For the
general problem of open string tachyon scattering, others \klebanov\
have found logarithmic divergences, but here we are looking only at
exceptional values of momenta for which their integrals are ill-defined.
One of the things we have accomplished here is to provide a
regularization scheme for which calculations at these usually singular
values of momenta are possible.

A careful analysis of the graphs shows that for the case at hand,
the graphs for the free energy with total charge $0$
(i.e. $\left|\sum q^X_j\right| + \left|\sum q^Y_j\right| = 0$)
diverge at most as $1/\epsilon$ while all other connected graphs are
finite. Naive power-counting would predict that the charge-one graphs
should diverge logarithmically (which is equivalent to saying that the
$\beta$-function for $e^{iX(t)}$ does not vanish).  Quite the contrary,
we have proved here that the $\beta$-function for $e^{iX(t)}$
vanishes at these special points.  In other work, we have also shown
that these theories exactly satisfy the infinite set of Ward identities
predicted by the connection with string theory outlined in Section 2.
An example is included in Appendix B, but the details of this
statement must await another paper.

This general analysis cannot say very much about the points in the phase
diagram where $\bab\notin Z$, but in Section 5 we will offer some
evidence in support of the claim, suggested by the results of our
renormalization group calculation in Section 3.2, that the there is a
{\it circle} of critical theories in the phase diagram to all orders in
the periodic potential.

\newsec{The Free Energy and Correlation Functions for Zero Magnetic Field}

We will now use the methods of the previous section to explicitly calculate the
free energy and correlation functions at the critical point with
$\beta=0$ and $\alpha = 1$. The mobility, or two-point function, at this
critical point has been calculated by  Fisher and Zwerger \fisher\ for a weak
potential and by Guinea {\it et al.} \ghm\ for the tight-binding model. We
provide a new calculation here for the weak potential
case which is in agreement with the previous ones, and extend it to all
$N$-point functions. Our method has the advantage that it is free of
approximation and works at some other points on the critical
circle with non-zero magnetic field. In addition, it can be used to calculate
correlation functions in the charged sector, which are of interest in
determining the renormalization-group flow and can be used to calculate
the boundary state in open string theory.

Even though we can perform any integral in the perturbation series for the
free energy and correlation functions, the one draw-back with our choice of
regulator is that calculations with it become quite time-consuming at higher
orders in $V_0$. To the order in $V_0$ that we have calculated,
the main features of the $\beta=0$ neutral sector which
emerge are that: the free energy is proportional to
$1/\epsilon+O(\epsilon)$ (the significance of this will be explained
Section 4.1); $\langle\dot X(t)\dot X(0)\rangle$
is proportional to its value at $V_0=0$; all higher $m$-point functions
of $\dot X$ are zero except for possible contact interactions;
and in the limit $\epsilon\to0$ all the correlation functions are $SU(1,1)$
invariant. For both the neutral and charged graphs, we find that the
only needed renormalizations are the subtraction of an infinite constant
from the free energy and a rescaling of the periodic potential strength.
For our choice of potential, this rescaling is equivalent to holding
$V_0 T\epsilon$ finite.  We will now describe some
of the calculations which lead to these conclusions.

\subsec{The Free Energy}

If we follow the steps described in Section 3 to introduce convenient
ultraviolet and infrared cutoffs, the partition function \Ztwon\ can be
rewritten as
\eqn\Zdef{Z=\sum_n{1\over(2n)!}
    \left[\left(V_0T\over 2\right)^2(e^\epsilon+e^{-\epsilon}-2)\right]^n
    {2n\choose n}Q_{2n}~,}
where $Q_{2n}$ is the integral defined in \Qtwonex.
The factor $e^\epsilon+e^{-\epsilon}-2$
comes from self-contractions of $X(t)$ with itself within a single
potential insertion $e^{\pm iX}$, and the factor of $T$ multiplying
$V_0$ comes from the change of variables from $t_i$ to $z_i$.
Performing the integrals and expanding in
powers of $\epsilon$, we have found that
\eqn\Qtcalc{Q_2={1\over e^\epsilon-e^{-\epsilon}}=
		{1\over2\epsilon}-{1\over12}\epsilon+
		{7\over720}\epsilon^3+O(\epsilon^5)~,}
\eqn\Qfcalc{Q_4={1\over2\epsilon^2}-{7\over32\epsilon}-{1\over6}+
                 {25\over192}\epsilon+O(\epsilon^2)~,}
and
\eqn\Qscalc{Q_6={3\over4\epsilon^3}-{63\over64\epsilon^2}
                +{7161\over65536\epsilon}+{3\over4}+O(\epsilon)~.}
Putting \Zdef\ through \Qscalc\ together, we can obtain the first few terms in
an expansion of the free energy, $F=-{1\over T}\ln Z$, in powers of the
potential strength:
\eqn\Ffinal{\eqalign{F=-{1\over T}
		&\left({V_0 T\epsilon\over2}\right)^2
             \left({1\over2\epsilon}+O(\epsilon)\right)
             -{1\over T}\left({V_0 T\epsilon\over2}\right)^4
            \left(-{7\over128\epsilon}+O(\epsilon)\right)\cr
            &-{1\over T}\left({V_0 T\epsilon\over2}\right)^6
            \left({10579\over786432\epsilon}+O(\epsilon)
		\right)+~\cdots~.}}
Since $\epsilon$ is the dimensionless ratio of the ultraviolet cutoff $\delta$
and the infrared cutoff $T$, it is useful to define a dimensionless
potential strength, $V_r=V_0\delta=V_0T\epsilon$, which makes no reference to
the infrared scale $T$. One of our main points will be that the large-$T$ limit
defines a critical theory, independent of the short-distance scale $\delta$,
once we have rescaled the potential strength in this way. The free energy is
then a power series in $\epsilon$ with coefficients which are themselves power
series in $V_r$:
\eqn\Frenorm{F =-{1\over T}\sum_{n=-1}^{\infty} f_n(V_r)
\epsilon^n~~.}
In the case at hand, the $f_0$ term happens to vanish, but we do not expect
this to be a general feature of the free energy.

It is instructive to re-express the free energy expansion in terms of the
dimensional scales $\delta$ and $T$:
\eqn\Fdens{F= -f_{-1}(V_r){1\over\delta} - f_0(V_r){1\over T}
		- f_{1}(V_r){\delta\over T^2} +~\cdots ~~~.}
The only universal term is the $f_0$ term: all the others
depend on $\delta$ and hence on the details of the definition of the theory at
short distances. Furthermore, the $f_{-1}$ term can be completely removed
by an appropriate $V_r$-dependent constant shift of the original
periodic potential (such a shift obviously has no observable effect on the
particle dynamics). The meaningful critical physics of the free energy is
therefore entirely contained in $f_0(V_r)$: In the
string theory context, it provides the overall normalization of the boundary
state (see \clny\ for examples). There is also a more conventional
thermodynamic
interpretation: In setting up our functional integral in Section 2, instead of
integrating out the oscillators, we could have shown that \parcon\ makes the
oscillators equivalent to a free massless scalar field living on a
one-dimensional world with a boundary (basically a spatial slice of the string
worldsheet) and with some non-trivial, but perfectly local, boundary action for
the massless scalar field induced by the action of the original quantum
particle. The path integral over paths periodic in Euclidean time with period
$T$ then generates the partition function of this system at temperature $\tau =
1/T$.  Equation \Fdens\ can then be interpreted as the expansion about zero
temperature  of the thermodynamic free energy and $f_0$ can be identified as
the zero-temperature limit of the entropy associated with the boundary.
(Affleck and Ludwig \afflud\ have recently calculated the boundary entropy of
Ising and Kondo systems, and their paper gives a clear explanation of how to
separate boundary from bulk contributions to the entropy.) In some sense, this
entropy counts the number of dynamically active degrees of freedom living on
the boundary. It should (and does) vanish for dynamically trivial boundary
conditions like Dirichlet or Neumann. What it should do for the case at hand,
where there is a complicated boundary action, is not obvious. The calculations
we have summarized in \Ffinal\ show, perhaps surprisingly, that $f_0(V_r)$ is
identically zero. We will later see that, if the magnetic field is turned on as
well as the potential, $f_0$ becomes a non-zero function of the magnetic field
and of $V_r$. In view of this discussion, in the rest of this paper we shall
identify the free energy with the $f_0$ term in the appropriate analog of the
expansion \Fdens. It would be more accurate to speak of the zero-temperature
entropy, but this abuse of language should cause no confusion.

\subsec{The $N$-Point Functions}

Next, we turn our attention to the $m$-point functions. When $m$
is odd, the $m$-point functions are zero by symmetry.  For the connected
$2m$-point functions, we insert the regulated version of \twoptin\ into the
partition function, and absorb the self
contractions of the potential insertions into $V_r$.  The result is
that we must compute the connected part of
\eqn\mptdef{\lb\dot X(r_1)\dots\dot X(r_{2m})\rb
            ={\lb\dot X(r_1)\dots\dot X(r_{2m})\rb}_0
             +\sum_{n=1}^\infty{(V_r/2)^{2n}\over(2n)!}
              {2n\choose n}\left({2\pi\over iT}\right)^{2m}A_{2n}~.}
Here, ${\langle\dot X(r_1)\dots\dot X(r_{2m})\rangle}_0$ is the $2m$-point
function in the absence of the periodic potential and $A_{2n}$ is defined to be
\eqn\Atndef{A_{2n}=\oint\prod_{j=1}^n{dz_j\over2\pi i}{dw_j\over2\pi i}
                     I_{2n}R_{n,m}~,}
where $I_{2n}$ is the function defined in \Itndef\ and
\eqn\Rnmdef{R_{n,m}=(-1)^m\prod_{j=1}^{2m}\sum_{i=1}^n
       \left({\xi_j^2-z_i^2\over(\xi_j-e^{-\epsilon}z_i)(\xi_j-e^\epsilon z_i)}
    -{\xi_j^2-w_i^2\over(\xi_j-e^{-\epsilon}w_i)(\xi_j-e^\epsilon
w_i)}\right),}
with $\xi_j=e^{2\pi ir_j/T}$.

For the two-point function, we have done explicit calculations out to fourth
order in the potential, finding
\eqn\Atcal{A_2=
     -{1\over2\sin^2\left[{2\pi\over T}\left(t_1-t_2\over2\right)\right]}
      + O(\epsilon^2)~;}
and
\eqn\Afcalc{A_4=\left(-{2\over\epsilon}+3\right)
		{1\over2\sin^2\left[{2\pi\over T}
		\left(t_1-t_2\over2\right)\right]}~.}
These two expressions are both proportional to the two-point function in the
absence of the periodic potential, which is given by
\eqn\Azdef{\lb\dot X(t_1)\dot X(t_2)\rb_0
           =-\left({2\pi\over T}\right)^2
   {1\over2\sin^2\left[{2\pi\over T}\left(t_1-t_2\over2\right)\right]}~.}
To this order in $V_0$, $A_{2n}$ contains one non-zero disconnected graph,
coming from $Z_2\cdot A_2$.  When we substitute the expression for $A_0$,
$A_2$ and $A_4$ back into equation \mptdef\ for the $2m$-point function, and
then subtract off the disconnected graph, we find that as $\epsilon\to0$,
\eqn\finaltpt{\lb\dot X(t_1)\dot X(t_2)\rb_{\rm conn}
            =\mu(V_r){\lb\dot X(t_1)\dot X(t_2)\rb}_0~,}
where $\mu$ is given by
\eqn\mudef{\mu(V_r)=1-\left({V_r\over 2}\right)^2
                    +{3\over4}\left({V_r\over2}\right)^4+~\cdots~~~.}
Therefore, the renormalized two-point function is equal to the ``free''
two-point function, rescaled by $\mu(V_r)$. For small enough $V_0$, we
see that $\mu<1$ but, lacking an all-orders calculation, we cannot show
that $\mu$ is always less than $1$.
(In ref. \cgcdef, we sketch an alternate calculation in which we
fermionize the theory. In that case, we can calculate the mobility to all
orders in $V_r$ and we find that $0\le\mu\le 1$.) We note here
that, after rescaling $V_0$, the answer is finite (except when $t_1=t_2$).
This means that we do not need to renormalize the kinetic part \SKEdef\
of the action, despite the fact that the
propagator correction diagrams appeared to have a divergence
$\propto1/\epsilon$.  These results are consistent with what was
found in the tight-binding approximation some time ago \ghm .

We have also used {\it Mathematica} to evaluate the integrals for the
four-point function to order $V_0^4$ and have found that the connected
four-point function, $\langle\dot X(r_1)\dots\dot X(r_{4})\rangle$, is zero
as $\epsilon\to0$, except when $r_i=r_j$.  In addition, we have
calculated the connected
$2m$-point functions for any $m>1$ at order $V_0^2$ and found them to
vanish as $\epsilon\to0$, as long as none of the points are coincident.
Both calculations are quite involved and the details of the latter are
given in Appendix A.

In summary, the critical theory looks basically like the free theory, except
that the two-point function is rescaled and the $2m$-point functions have
contact terms. The critical theory is obtained as the large-$T$
(zero-temperature!) limit of the theory with a fixed short-distance scale. The
only needed renormalizations are a rescaling of the potential strength (the
critical theory depends only on $V_r=V_0\delta$) and the subtraction of
a temperature-independent constant from $F$. The two-point function differs
from the free two-point function by a finite mobility factor $\mu(V_r)$ which
decreases from unity as $V_r$ increases from zero.  We expect that the mobility
should be less than one for any $V_r$  because the periodic potential
should inhibit the  particle's motion. Lastly, the critical theory is
$SU(1,1)$-invariant: The interacting two-point function is proportional to the
free two-point function, which is $SU(1,1)$-invariant, and the higher-point
functions, barring contact terms, are zero. (Actually, using the fermionic
regulator of \cgcdef, we can show that even the contact terms are
$SU(1,1)$-invariant.)

\newsec{Magnetic Field Effects}

We now turn to our main problem, that of calculating the critical
properties of a particle moving in two dimensions and simultaneously subject to
a magnetic field and a periodic potential. As before, we will calculate
both the free energy and a variety of $N$-point functions in a
perturbation expansion in the potential strength (but the dependence on
the magnetic field will be exactly accounted for). We will also restrict
ourselves to the neutral sector of the Coulomb gas arising from the
expansion in powers of the periodic potential. In what follows we assume
that we are on the critical circle $\aab=1$ and identify where we are
on that circle by the value of $\gamma=\beta/\alpha$.
So far we have proved to all orders in $V_0$ that the points
where $\gamma\in Z$ are exactly critical, and have shown in Section
3.2 that to the first few orders in $V_0$ the theory is also critical for
$\gamma\notin Z$.  We now support this with some explicit calculations,
valid for any $\gamma$, which again show no logarithmic divergences.

\subsec{Magnetic Field Contribution to Free Energy}

Let us first deal with the free energy, expanding it in powers of the
potential strength: $F(B,V_r)=-\sum_0^\infty F_n(B)V_r^n$. The term of
zeroth-order in $V_r$ has been computed elsewhere \refs{\tseyt,\abou}
and has the interesting form
\eqn\Binfeld{F_0(B) = {1\over2T}\ln(1 +(2\pi\apm B)^2)~.}
As explained in \abou\ and elsewhere, if this expression is treated as
an action functional, it generates stringy corrections to Maxwell's
equations. It is at least an existence proof that the boundary entropy
discussed in the previous section can have non-trivial dependence on the
parameters of the critical theory. We will not discuss it further, as we are
really interested in the joint effects of potential and magnetic field which
are responsible for the unusual spectrum of the Hofstadter problem with no
dissipation. Such effects do not occur until fourth-order in the expansion in
powers of the periodic potential. This is the first order at which we can have
the insertion of both $X$ and $Y$ ``charges'', so that the system can feel the
effect of the phases due to the magnetic field.  (By the zero-charge condition,
each $e^{iX}$ must be accompanied by an $e^{-iX}$ and  each $e^{iY}$ by an
$e^{-iY}$.)  In what follows, we will evaluate the free energy to order $V_0^4$
(it is possible, but tedious to go to higher orders).

For calculations involving non-zero magnetic field, the contour
integral technique introduced in Section 3 encounters difficulties at
fourth order in the potential strength.  This is precisely because
the interesting interplay between $B$ and $V$ begins at this order, and
the origin of the problem can be seen explicitly in the part of \eGmn\
which is the exponential of the off-diagonal propagator.  Diagrams with
both $X$ and $Y$ internal charges will involve integrals over expressions
containing factors of this kind, and whenever $\gamma$ is not an integer,
branch cuts will be present which make evaluation difficult.  To avoid
this complication, we choose to use a new method of calculation starting
directly with equation \ZnBln\ for the partition function.  In the
unregulated diagrams, the exponentiated off-diagonal
propagators which connect $X$ and $Y$ charges simply contribute a
phase factor when an $X$ charge moves past a $Y$ charge.
Except for the above-mentioned $V_0$-independent piece,
the magnetic field only contributes to the free energy through these
phases.  Thus we will take into account the interaction between the
the $X$ and $Y$ charges simply by keeping track of the phase factors.
We will regulate the diagonal propagator in the usual way.
Unfortunately, the necessity to maintain the ordering of the charges
makes it impossible to use contour integration and so we
resort to series expansion of the diagonal propagator instead.
The advantage of this technique of course is that it works for
non-integer values of $\gamma$.

One might worry that we are not regulating the $\sign$-function which
generates the phases, but this is only of concern in the calculation of
correlation functions.  If we simply wish to calculate the free energy,
the off-diagonal propagator contributes no divergences, and we expect
that the phase prescription will work correctly.  In calculating
correlation functions, however, we take derivatives of the off-diagonal
propagator and regulation becomes necessary because the derivative of
a $\sign$-function is a $\delta$-function.

As we indicated above, the first $B$-dependence in the free energy
comes at order $V_0^4$, and we now proceed to calculate this term.
The contribution to the partition function of the diagram with $n_X$ of
the $X$ charges and $n_Y$ of the $Y$ charges is (with $n=n_X+n_Y$)
\eqn\Znwithb{\eqalign{Z(n_X,n_Y)= &\left({V_0\over2}
		e^{-\shalf G(0)}\right)^n\cr
                   &\sum_{\vec q_i ={(\pm 1,0)\atop(0, \pm1)}}
                    \int d\tau_1 \dots d\tau_n ~
                    \exp\sum_{i<j}\bigl(i\phi_{ij}
	-\vec{q}_i\cdot\vec{q}_j G(\tau_i-\tau_j)\bigr)}}
where the phase factor $\phi_{ij}$ is defined by
\eqn\phase{\phi_{ij}=\pi\gamma\epsilon^{\mu\nu}q_i^\mu q_j^\nu
			\sign(\tau_i-\tau_j) ~,}
the integration range is $0\le\tau_i\le\tau_{i+1}\le T$ and $G(t)$ is the
diagonal element of the propagator \Gtdef.  The fourth-order piece of the
free energy is given by
\eqn\free{\eqalign{F_4=-{4\over T}\left({V_0\over2}e^{-\shalf G(0)}\right)^4
              & \left(e^{2\pi i\gamma}+e^{-2\pi i\gamma}-2\right)\cr
               &\int dt_1dt_2ds_1ds_2 ~
               \exp\bigl(G(t_1-t_2)+G(s_1-s_2)\bigr)}}
where: $0\le t_1\le s_1\le t_2\le s_2\le T$; the overall factor of 4 takes
account of the fact that the first charge can be a plus or a minus and an $X$
or
a $Y$; the $-2$ in the $\gamma$-dependent factor subtracts the contribution
due to the disconnected graphs, which are also present in the absence
of a magnetic field.

To get finite results we must, of course, regulate the propagator. Because we
are on the critical circle, we can use the strategy described in Section 3
following \Gmmdef. This amounts to the replacement
\eqn\prop{G(z)=-\ln\left(1-ze^{-\epsilon}\right)
                   \left(1-\bar{z}e^{-\epsilon}\right)}
with $z=\exp\left(2\pi i t/T\right)$.
For our later calculation it will be useful to have the following power series
expansion of the exponentiated propagator
\eqn\exprop{e^{G(z)}={1\over{1-e^{-2\epsilon}}}\sum_{m=-\infty}^{\infty}
                z^m e^{-\left|m\right|\epsilon}~.}
For coincident points this reduces to
\eqn\expgz{e^{-G(0)}=\left(1-e^{-\epsilon}\right)^2~.}
At this point we note the translational invariance and periodicity of
$G$ and use the formula
\eqn\perint{\int_0^T{d\tau_1\dots d\tau_n\over{n!}}f(\tau_i-\tau_j)
            ={T\over{n}}\left.\int d\tau_2\dots d\tau_n f(\tau_i-\tau_j)
             \right|_{\tau_1=0}}
where on the right hand side $0\le\tau_i\le\tau_{i+1}\le T$ and $f(\tau)$
is periodic in all its arguments with period $T$.  This allows us to write
\eqn\freei{F_4=4\sin^2\pi\gamma\left({V_0\over2}e^{-\shalf G(0)}\right)^4
             \int ds_1dt_2ds_2 \exp\bigl(G(t_2)+G(s_1-s_2)\bigr)}
where again $0\le s_1\le t_2\le s_2\le T$.  Transforming to angular
variables on a circle and using \exprop\ we obtain
\eqn\freeii{F_4=4\sin^2\pi\gamma\left({V_0\over2}e^{-\shalf G(0)}\right)^4
              \left({T\over2\pi}\right)^3
              \left({1\over1-e^{-2\epsilon}}\right)^2
              \sum_{m,n=-\infty}^{\infty}
              e^{-\left|m\right|\epsilon}e^{-\left|n\right|\epsilon}
              I_{mn}}
where the integral is given by
\eqn\imn{I_{mn}=\int_0^{2\pi}d\phi_2\int_0^{\phi_2}d\theta_2
                \int_0^{\theta_2}d\phi_1
                e^{im\theta_2}e^{in(\phi_2-\phi_1)}~.}
Doing the integral we find
\eqn\imnans{\eqalign{I_{mn}={4\pi^3\over3}\delta_{m,0}&\delta_{n,0}
                -{4\pi\over m^2}\delta_{n,0}\left(1-\delta_{m,0}\right)
                -{4\pi\over n^2}\delta_{m,0}\left(1-\delta_{n,0}\right)\cr
                +&{2\pi\over n^2}\left(1-\delta_{n,0}\right)
                \left(\delta_{m,n}+\delta_{m,-n}\right)\cr}}
and therefore that
\eqn\sumimn{\sum_{m,n}e^{-\left|m\right|\epsilon}
            e^{-\left|n\right|\epsilon} I_{mn}= {4\pi^3\over3}
            -8\pi\sum_{m=1}^{\infty}{1\over m^2}
		\bigl(2e^{-m\epsilon}-e^{-2m\epsilon}\bigr)~.}
The sum can be evaluated for small $\epsilon$ and we find that
\eqn\sumans{\sum_{m=1}^{\infty}{1\over m^2}
		\bigl(2e^{-m\epsilon}-e^{-2m\epsilon}\bigr)
		={\pi^2\over6}-2\epsilon\ln2+{\epsilon^2\over2}
		-{\epsilon^3\over12}+O(\epsilon^4)~.}
Finally, using the renormalization of the potential strength
introduced in the previous section, the free energy becomes
\eqn\fans{F_4={V_r^4\over32\pi^2T}\sin^2\pi\gamma
            \left({4\ln2\over\epsilon}-1+O(\epsilon)\right)~.}
The $\epsilon^{-1}$ term must of course be subtracted away, but a finite
magnetic-field-dependent piece is left over. This finite part vanishes
for any integer $\gamma=k$. This is physically reasonable since, at these
points, the magnetic phase associated with transporting the
center of the electron's orbit around a unit cell of the reciprocal
lattice (as defined in equation \reclat) is a multiple of $2\pi$.
Therefore we expect this situation to look equivalent to the zero field case.
Consideration of the phase prescription for higher orders in the
potential shows that all such terms will also be zero for these
special points on the critical circle.  Of particular importance is the
observation that there is no logarithmic divergence even when $\gamma$
is not an integer.  If this remains true for higher orders,
as we believe it will, then it confirms the result of Section
3.2 that we have a critical {\it circle}, $\alpha=\alpha^2+\beta^2$.

\subsec{$N$-Point Functions and Magnetic Field}

We now turn to the calculation of the $m$-point functions on the
critical circle.  Again, the $(2m+1)$-point functions are all trivially
zero by symmetry arguments.  We present explicit calculations of the
$2m$-point functions to $O(V_0^2)$ using the contour integral techniques
outlined in Section 3. Our results are valid for any value of $\gamma$,
not just integer ones.  As we saw above, the extension of these results
to higher orders in the strength of the periodic potential is more
troublesome because, unless $\gamma$ is an integer, there are branch
cuts in the integrands whenever there is an internal line connecting
internal $X$ and $Y$ charges.  However, our earlier calculations show
that when $\gamma$ is an integer, nothing remarkable happens at higher
orders, and certainly that there will be no logarithmic divergences.
For non-integer $\gamma$, the presence of the branch cuts
makes it harder to be certain, but the absence of logarithmic
divergences in the free energy is strong evidence that even here the
$O(V_0^2)$ behavior will carry on essentially unchanged at higher
orders.

The $O(V_0^2)$ contribution to the $2m$-point functions has the form
\eqn\magcordef{\eqalign{\lb\dot X^{\mu_1}(r_1)\dots
	\dot X^{\mu_{2m}}(r_{2m})\rb_2
	= ~ & \cr
	\left({V_0\over2}e^{-\shalf G(0)}\right)^2
	\int dt_1 dt_2&~\exp\bigl(G^{XX}(t_1-t_2)\bigr)
	\left[R^X(2m) + R^Y(2m)\right]~,}}
where
\eqn\magRdef{R^\nu(2m)=(-1)^m\prod_{i=1}^{2m}
      \left[{d\over dr_i}G^{\mu_i\nu}(r_i-t_1)
       -{d\over dr_i}G^{\mu_i\nu}(r_i-t_2)\right].}
$R^X(2m)$ comes from two $\cos X$ potential insertions, and $R^Y(2m)$
comes from two $\cos Y$ potential insertions.   Now let $G$ be
equal to $G^{XX}$ and $N$ equal to $G^{XY}$, where the propagators have
been regulated as in Section 3.  Then, with the definition
$z_j=\exp(2\pi i t_j/T)$, we have
\eqn\gdef{\del_1 G(t_1-t_2)={2\pi i\over T}\left[
			{z_2\over z_2-e^{-\epsilon}z_1}
			-{z_1\over z_1-e^{-\epsilon}z_2}\right]}
and
\eqn\ndef{\del_1 N(t_1-t_2)=\gamma{2\pi i\over T}
		\left[{z_1\over z_1-e^\epsilon z_2}
			+{z_2\over z_2-e^\epsilon z_1}\right].}

We will first restrict our attention to the calculation of the
$\langle\dot{X}\dot{X}\rangle$ and $\langle\dot{X}\dot{Y}\rangle$
two-point functions.  Because our system is invariant under rotations in the
$X$-$Y$ plane,
$\langle\dot{Y}\dot{Y}\rangle$ has the same form as
$\langle\dot{X}\dot{X}\rangle$.  The integrals for the second order
contributions to the two-point functions can be written as
\eqn\diag{\lb\dot{X}(t_1)\dot{X}(t_2)\rb_2 =
		-\left({V_0\over2}e^{-\shalf G(0)}\right)^2
		\int_0^T du dv~e^{G(u-v)} I_\alpha (u,v)}
and
\eqn\difodiag{\lb\dot{X}(t_1)\dot{Y}(t_2)\rb_2 =
		-\left({V_0\over2}e^{-\shalf G(0)}\right)^2
		\int_0^T du dv~e^{G(u-v)} I_\beta (u,v)}
where $I_\alpha$ is defined by
\eqn\iddef{\eqalign{I_\alpha (u,v)=
	2\del_1G(t_1-u)\del_2\bigl(&G(t_2-u)-G(t_2-v)\bigr)\cr
	&+2\del_1N(t_1-u)\del_2\bigl(N(t_2-u)-N(t_2-v)\bigr),}}
and $I_\beta$ is given by
\eqn\ioddef{\eqalign{I_\beta (u,v)=
	2\del_1G(t_1-u)\del_2\bigl(&N(t_2-v)-N(t_2-u)\bigr)\cr
	&+2\del_1N(t_1-u)\del_2\bigl(G(t_2-u)-G(t_2-v)\bigr).}}
The calculation by contour integration is straight-forward and gives for
the diagonal contribution
\eqn\diagans{\eqalign{\lb\dot{X}(t_1)\dot{X}(t_2)\rb_2 =&
             \left(V_r\over 2\right)^2
             \left({2\pi\over T}\right)^2 (1-\gamma^2)
             \left[2\sin^2\left({\pi\over T}
             (t_1-t_2)\right)\right]^{-1} \cr
             =&-(1-\gamma^2)\left(V_r\over 2\right)^2
             \lb\dot{X}(t_1)\dot{X}(t_2)\rb_0}}
where $V_r$ is the renormalized potential, given by $V_r = V_0T\epsilon$.
This two-point function is the same as the free propagator with a rescaled
coefficient depending on the renormalized potential strength and the
magnetic field.

For the off-diagonal part we get zero for separated points
and the contact term looks like the derivative of a delta function, or
two derivatives of a step function, as we would expect if this were also
just a rescaled version of the free off-diagonal propagator.
Explicitly, we find
\eqn\difodans{\lb\dot{X}(t_1)\dot{Y}(t_2)\rb_2 =
		-i\pi\gamma V_r^2\delta'(t_1-t_2)~.}
Integration of the regulated version of this result, or a direct
calculation using the series expansion \exprop\ for $G$ (and a similar
one for $N$) gives the result that
\eqn\odans{\eqalign{\lb X(t_1)Y(t_2)\rb_2  &=
		-2\pi i\gamma\left({V_r\over2}\right)^2
		\left({2\over T}(t_1-t_2)-\sign(t_1-t_2)
		\right) \cr
			&= -2\left({V_r\over2}\right)^2
		\lb X(t_1) Y(t_2)\rb_0~.}}
The constants of integration which arise upon integrating \difodans\
twice to obtain \odans\ are fixed by the requirement that all functions
be periodic with period $T$.  Thus the second order contribution is just
$-2(V_r/2)^2$ times the free propagator.
The conclusion here is that when we turn on the potential in the
presence of the magnetic field, the zeroth-order two-point function
$\langle\dot{X}^\mu\dot{X}^\nu\rangle$ is just multiplied by a constant
matrix whose elements are functions of $\gamma$ and $V_r$.
For the special points in the phase diagram where
$\gamma$ is integer we know for sure that this remains true at higher
orders in the potential, but we have not proved this when $\gamma$ is
not an integer.

In reference \cgcdef, we have shown that the correlation functions on the
critical circle at integer $\gamma$ are all related by a duality symmetry.
Given the form of the two-point function for $\gamma=1$, this symmetry
completely determines the form of the two-point functions at all the
other special critical points.  It is interesting to compare the
duality prediction with the explicit functions we have just calculated.
To do this, it is convenient to specify location in the $\alpha$-$\beta$
plane by the complex number $z = \alpha + i \beta$ and to rewrite the
free two-point function \Gtdef\ as follows:
\eqn\Dtdef{G^{\mu\nu}(t;z)=-{\rm Re}(1/z)~\ln(t^2)~\delta^{\mu\nu}
                  + i\pi~{\rm Im}(1/z)~\sign(t)~\epsilon^{\mu\nu}.}
(For simplicity we have removed the infrared cutoff and rewritten this
as a two-point function on the open line.)
The content of the duality relation derived in \cgcdef\ is that we can
express the interacting two-point function at the special points on the
critical circle solely in terms of the function $G$ at various values of
$z$, and of the zero-magnetic-field mobility function, $\mu(V_r)$,
displayed in \mudef.  Define $z_\gamma= 1/(1-i\gamma)$.
Then when $\alpha/(\alpha^2 + \beta^2) = 1$ and $\beta/\alpha = \gamma$,
we have $z$ equal to $z_\gamma$.
According to equation 6.2 in \cgcdef, when $\gamma$ is an integer, the
two-point function with potential strength $V_0$ is given by
\eqn\dualdef{\lb\vX(t)\vX(0)\rb(\gamma,V_0)
     =G(t;z_\gamma) + \left(\mu(V_r)-1\right)G(t;{z_\gamma}^2)~.}
The first term on the right-hand side is just the expression for the two-point
function in the absence of the potential.  The second term contains all the
effects due to the potential.  We note that
$1/{z_\gamma}^2 = (1-\gamma^2) - 2i\gamma$, and that
$\left[\mu(V_r)-1\right] = -\left(V_r/2\right)^2 + O(V_r^4)$.
Putting this together with \Dtdef, we see that \dualdef\ predicts
the order $V_0^2$ part of the two-point function to be
\eqn\dualtpt{\eqalign{\lb X^{\mu}(t)X^{\nu}(0)\rb(\gamma,V_0)
        =\left({V_r\over 2}\right)^2
           \left(1-\gamma^2\right)&\ln(t^2)~\delta^{\mu\nu}\cr
         &+i\pi\left({V_r\over 2} \right)^2 2~\gamma~
         \sign(t)~\epsilon^{\mu\nu}.}}
This can be expressed in terms of the free two-point function \Dtdef\
at $\gamma$ as follows:
\eqn\dualxx{\lb \dot X(t) \dot X(0)\rb_2 (\gamma, V_0)
    = -(1-\gamma^2)\left({V_r\over 2}\right)^2
      {\lb\dot X(t) \dot X(0) \rb}_0~,}
and
\eqn\dualxy{\lb \dot X(t) \dot Y(0)\rb_2 (\gamma, V_0)
    = -2\left({V_r\over 2}\right)^2
      {\lb\dot X(t) \dot Y(0) \rb}_0~.}
On comparing this expression with equations \diagans\ and \odans, we see that
it agrees with our direct computation.  We conclude that not only do the
two-point functions exhibit $SL(2,R)$ covariance as a function of the
time variable, but they also satisfy a duality symmetry that relates
two-point functions at different values of friction and flux.

For the $2m$-point functions with $m\ge2$, we can again perform the contour
integrals and sum over all the diagrams, exactly as we do in Appendix A for
the $2m$-point functions with zero magnetic field.  This time, we find that
not all the correlation functions vanish when the points are non-coincident.
To this order in the potential, any correlation function with exactly two
$\dot X$'s or two $\dot Y$'s is finite as the cut-off goes to zero and
all the other correlation functions are zero except for
contact terms.  When there are exactly two $\dot X$'s, the result is
\eqn\magcorfin{
    \lb\dot X(t_1)\dot X(t_2)\dot Y(s_1)\dots \dot Y (s_{2m})\rb_2=
     C_m \prod_{j=1}^{2m}
       \left({z_1+w_j\over z_1-w_j}-{z_2+w_j\over z_2-w_j}\right)
       {z_1 z_2\over (z_1-z_2)^2}~,}
for $m\ge2$, and similarly when there are exactly two $\dot Y$'s.  In this
equation, we have defined
\eqn\cdef{C_m = \left({2\pi\over T}\right)^{(2m+2)}
		\left({V_r\over 2}\right)^2 \gamma^2~,}
$z_j = e^{2\pi i t_j/T}$ and $w_j = e^{2\pi i s_j/T}$.
The four-point function can be written more simply as
\eqn\bfourpt{\lb\dot X(t_1)\dot X(t_2)\dot Y(s_1)\dot Y(s_2)\rb_2
  = C_1 {16z_1 z_2 w_1 w_2\over (z_1-w_1)(w_1-z_2)(z_2-w_2)(w_2-z_2)}~.}
We note that these correlation functions have the same form as the integrand
of the $O(V_0^2)$ contribution of the $2m$-point function with no magnetic
field.  Using arguments similar to those in Section 3, we can show that
such functions are $SU(1,1)$ covariant (or $SL(2,R)$ covariant in the limit
as $T\to\infty$).  Thus, we have found non-trivial critical theories which
exhibit not only scale invariance, but also the higher symmetry group,
$SL(2,R)$, as predicted by their connection with string theory.
At higher orders in $V_0$, when $\gamma$ is an integer,
we expect that additional correlation functions will have a finite,
$SL(2,R)$-covariant limit as the cut-off is taken to zero.


\newsec{Charged Sector Diagrams}

So far in this paper we have restricted our attention to the
{\it neutral} sector of the Coulomb gas arising from the perturbative
expansion in the periodic potential. In the DQM path integral, this
restriction is enforced by the integral over the zero-mode. In string theory,
the zero-mode integral usually serves to enforce momentum conservation in
S-matrix elements. However, in calculating the open string boundary state,
the zero-mode integration is left undone, and the charged sector contributes.
Charged diagrams are also of interest in their own right from a Coulomb
gas point of view and are useful in calculating the renormalization-group
flow ({\it e.g.} as in Section 3.2 or in \klebanov).
We shall see that in the critical theory,
the connected diagrams of the charged sectors are all completely finite
functions of the rescaled potential $V_r$. For simplicity,
we restrict ourselves in this section to zero magnetic field.

\subsec{Charged Sector Free Energy}

As shown previously for the critical theory at zero magnetic field,
all diagrams can be calculated by contour integration tricks.
The $n$th order contribution to the partition function is given by
\eqn\pfn{Z_n = {1\over n!}\left({V_0\over2}e^{-\shalf G(0)}\right)^n
		\sum_{\{q_j=\pm1\}} e^{i Q_n X_0}
		\int_{-T/2}^{T/2} \prod_{i=1}^n dt_i ~
		\exp\bigl(-\sum_{j<k} q_j q_k G(t_j-t_k)\bigr)}
where $Q_n=\sum q_j$, $G(t)$ is defined by eqn. \prop\ and
$X_0$ is the zero-mode.  We have done the calculation to fourth order
in $V_0$ and found the following results for $Z$ (we express
everything in terms of the usual rescaled potential strength
$V_r =V_0T\epsilon$):
\eqn\zeroth{Z_0=1~,}
\eqn\first{Z_1 =2\left({V_r\over2}\right)\cos X_0
		\left[1-{\epsilon\over2} + O(\epsilon^2)\right]~,}
\eqn\second{Z_2=2\left({V_r\over2}\right)^2
		\left[{1\over4\epsilon}+(1-2\epsilon)\cos2 X_0
                 -{\epsilon\over48} +O(\epsilon^2)\right]~,}
\eqn\third{Z_3 =2\left({V_r\over2}\right)^3
		\left[{1\over2\epsilon}\cos X_0
		+\cos3 X_0 -{5\over 8}\cos X_0
		+O(\epsilon)\right]~,}
\eqn\fourth{\eqalign{Z_4=2\left({V_r\over2}\right)^4 &
		\left[\left({1\over4\epsilon}\right)^2
		+{1\over2\epsilon}\left(\cos2X_0-{7\over 128}\right)
		\right.\cr &\left.\qquad
		-{1\over96} +\cos4X_0-{5\over3}\cos2X_0
		+O(\epsilon)\right]~.}}
The free energy is related to the connected part of these diagrams in the
usual way:
\eqn\free{F=-{1\over T}\ln Z =
	-{2\over T}\sum_{n=1}^\infty \left({V_r\over2}\right)^n F_n~~.}
Rearranging the series for $Z$, we find that the first four $F_n$ are
\eqn\Fi{F_1 = \cos X_0~,}
\eqn\Fii{F_2 = {1\over4\epsilon} +{1\over2}\cos 2X_0 -{1\over2}~,}
\eqn\Fiii{F_3 ={1\over3}\cos 3X_0 -{3\over8}\cos X_0~,}
\eqn\Fiv{F_4 = 	-{7\over256\epsilon} +{1\over4}\cos 4X_0
		-{7\over24}\cos 2X_0 +{1\over8}~.}
The contribution of the charge-Q diagrams is given by the coefficient of
$e^{iQX_0}$ in these expressions. All the divergences come from the neutral
sector and can be removed by the subtraction of
a $V_r$-dependent $1/\epsilon$ term.  Using general arguments about the
form of the diagrams, we can show that this behavior
continues to all orders.  The key thing to note here
is that the logarithmic divergences expected in such diagrams
according to naive power-counting arguments
are completely absent.  Using our method of calculation, this is of
course guaranteed from the beginning.

\subsec{Charged Sector Two-Point Function}

Calculation of the $N$-point functions can be done in a very similar way.
The two-point function is particularly important for the open string
boundary state (one extracts the energy-momentum tensor from it)
and also for checking consistency with the Ward identity, so we show
the calculation here as an example.  The explicit formula for the
$n$-th order contribution to the two-point function is
\eqn\ntwopt{\lb\dot X(\tau_1)\dot X(\tau_2)\rb_n=
		-{1\over Z}{1\over n!}
		\left({V_0\over2}e^{-\shalf G(0)}\right)^n
		\sum_{q_j=\pm1} e^{iQ_n X_0}
		\sum_{i,j=1}^n q_i q_j I_{ij}(n,\vec{q})}
where we define $I_{ij}(n,\vec{q})$ by
\eqn\inqdef{I_{ij}(n,\vec{q})=\int dt_1 \dots dt_n
		~\del_1 G(\tau_1-t_i)~\del_2 G(\tau_2-t_j)
		\prod_{k=2}^n \prod_{l=1}^{k-1}
		\exp\bigl(q_k q_l G(t_k-t_l)\bigr)~.}
Explicit results for the first three terms in an expansion in powers of
the potential are
\eqn\tpo{\lb\dot X(\tau_1)\dot X(\tau_2)\rb_0
         =-\left({2\pi\over T}\right)^2
           \left(2\sin^2\bigl({\pi\over T}
           \tau_{12}\bigr)\right)^{-1}~,}
\eqn\tpi{\lb\dot X(\tau_1)\dot X(\tau_2)\rb_1
         =V_r \cos X_0 \left({2\pi\over T}\right)^2
          \left(1-\Delta_{2\epsilon}
		\bigl({2\pi\over T}\tau_{12}\bigr)\right)~,}
\eqn\tpii{\eqalign{\lb\dot X(\tau_1)\dot X(\tau_2)\rb_2
           =\left({V_r\over2}\right)^2& \left({2\pi\over T}\right)^2
            \biggl[8\cos{2X_0}
            \cos^2\bigl({\pi\over T}\tau_{12}\bigr)
		-4\cos^2X_0\biggr.\cr
            &\biggl.+4\sin^2 X_0~\Delta_{2\epsilon}
		\bigl({2\pi\over T}\tau_{12}\bigr)
            +\left(2\sin^2\bigl({\pi\over T}\tau_{12}\bigr)\right)^{-1}
            \biggr]~,}}
where $\Delta$ is a regulated delta-function defined by
\eqn\ddef{\Delta_{n\epsilon}(\tau)=\lim_{\epsilon\rightarrow0}
          \left({\sinh(n\epsilon)\over\cosh(n\epsilon)-\cos\tau}\right)}
and $\tau_{12}=\tau_1-\tau_2$.  Note that while all divergences are
eliminated by rescaling the potential, the answer contains contact
terms.

An important aspect of this result is that the zero-mode-dependent pieces
are not $SU(1,1)$-invariant. This has to do with the fact, explained in
\CTrivi, that reparametrization invariance of the boundary state path
integral is not manifest: it is in some sense ``softly broken''
by the non-local
kinetic term in the boundary state action and the fixing of the zero mode
in the boundary state path integral measure. There is nonetheless a complete
set of ``broken reparametrization invariance'' Ward identities in which all
of these effects are included \CTrivi. As it turns out, the full Ward identity
for the zero-mode-dependent two-point function involves the zero-mode-dependent
part of the the free energy as well. The calculations reported in this section
can therefore be used to make a rather nontrivial check of the Ward identities
(not to speak of our whole renormalization scheme). We have checked that the
above expressions satisfy the identity. The presence of the contact terms
(terms involving $\Delta$) in the two-point function turns out to be a crucial
element in the consistency of the Ward identity (details are relegated to an
appendix). We conclude that our renormalization scheme produces a genuine
solution to open string theory.

Given these results we can also explicitly construct the tachyon and
massless particle contributions to the boundary
state, defined in equation \bs, up to second order in the tachyon
strength, $V_0$.  From the massless contribution, space-time
equations of motion
can be derived for the graviton and dilaton in the presence of this
background field, and the stress energy tensor in Einstein's equations
due to the tachyon source can be constructed explicitly.  However,
efforts to discover a space-time effective action from which the
equations of motion could be derived were unsuccessful, probably because
the tachyon fluctuates on a short scale, rendering meaningless a
low-energy description of its effects.

\newsec{Conclusions}

The state of affairs described in this paper is promising, but far from
fully satisfactory. Our goal is to find as complete a characterization of
the ``Hofstadter'' critical theories as one has for the Ising model
or the WZW models. The purely perturbative approach described in this
paper is obviously not going to take us that far, but it does give a pretty
clear idea of what we would eventually like to prove. Since one of
the marginal terms (the dissipation term) in the action is non-local, standard
field theory intuition does not necessarily apply and we have had to invent
special methods and re-examine the whole question of critical behavior from
the ground up. Our primary claim, supported by a variety of perturbative
calculations, is that there is a two-parameter critical surface: The starting
action has three marginal parameters (dissipation constant, magnetic
field strength and strength of the periodic potential) and we have
offered evidence that there is a critical theory for any value of the
renormalized potential strength so long as the dissipation constant and
magnetic field satisfy one functional condition.  (Basically the periodic
potential has to be a dimension-one operator in the one-loop approximation!
This is reminiscent of the situation in the Liouville theory.)
We have also calculated some of the critical $N$-point functions. Because we
are working in the rather unfamiliar context of critical theories with
nonlocal interactions, we have found it worthwhile to use these $N$-point
functions to explicitly check the reparametrization invariance Ward identities
(this ensures that the theory is not just critical, but also a solution
of string theory, a much more stringent requirement). A further, surprising,
finding is that many of the higher $N$-point functions
({\it all} of them in the case of zero magnetic field)
reduce to contact terms. This means that these
critical theories are almost, but not quite, trivial. This is a broad hint
that an exact solution for this system should be possible, but we have
yet to see how to exploit it. We hope to return to this point in a future
publication.
\vskip .3in
\centerline{\bf Acknowledgements}

This work was supported in part by DOE grant DE-AC02-84-1553.  D.F. was
also supported by DOE grant DE-AC02-76ER03069 and by NSF grant 87-08447.
A.F. also received support from an Upton Foundation Fellowship.

\appendix{A}{$N$-Point Functions}

In this appendix, for zero magnetic field, we evaluate the $O(V_0^2)$
contribution to the higher
$2m$-point functions of $\dot X$ and demonstrate that, except for contact
terms, they are zero as $\epsilon\to0$.  As described in Section 4, the
integral we want to evaluate is $A_2$ \Atndef:
\eqn\Atwodef{A_2= \oint{dz_1\over 2\pi i}{dz_2\over 2\pi i} I_2 R_{1,m}~,}
with
\eqn\Itwodef{I_2=-{1\over (z_1-e^{-\epsilon} z_2)(z_1-e^\epsilon z_2)}}
and
\eqn\Rmdef{\eqalign{R_{1,m}=&(-1)^m\prod_{j=1}^{2m}
       \left({w_j^2-z_1^2\over (w_j-e^{-\epsilon}z_1)(w_j-e^\epsilon z_1)}
      -{w_j^2-z_2^2\over (w_j-e^{-\epsilon}z_2)(w_j-e^\epsilon z_2)}\right)\cr
      =&(-1)^m\sum_{M=0}^{2m}\sum_{\sigma_M}\prod_{j\in\sigma_M}
        {w_j^2-z_1^2\over (w_j-e^{-\epsilon}z_1)(w_j-e^\epsilon z_1)}\cr
      &\qquad\qquad\times\prod_{i\notin\sigma_M}
      {w_i^2-z_2^2\over (w_i-e^{-\epsilon}z_2)(w_i-e^\epsilon z_2)}
       (-1)^M~,}}
where $\sigma_M$ is summed over all subsets of the first $2m$ integers which
contain $M$ elements.

We will proceed by performing the integral
\eqn\firstint{\oint{dz_1\over 2\pi i}{dz_2\over 2\pi i}
     \prod_{j=1}^M{w_j^2-z_1^2\over (w_j-e^{-\epsilon}z_1)(w_j-e^\epsilon z_1)}
     \prod_{i=M+1}^{2m}
      {w_i^2-z_2^2\over (w_i-e^{-\epsilon}z_2)(w_i-e^\epsilon z_2)} I_2}
and then symmetrize later.  If we first perform the $z_1$ integral, we must
evaluate the residues when $z_1=e^{-\epsilon}z_2$ and, if $M>0$, also when
$z_1=e^{-\epsilon}w_k$, for $1\le k\le M$.  We will call the residue from
the $z_1=e^{-\epsilon}z_2$ pole $r_z$, and the residues from the
$z_1=e^{-\epsilon} w_k$ poles $r_k$.  They are given by
\eqn\rzdef{r_z = {1\over z_2 (e^{\epsilon}-e^{-\epsilon})}
          \prod_{i=1}^M{(z_2^2e^{-\epsilon}-w_i^2e^\epsilon)\over
             (z_2-w_i)(z_2e^{-\epsilon}-w_i e^\epsilon)}
     \prod_{j=M+1}^{2m}
      {z_2^2-w_j^2\over (z_2-e^{-\epsilon}w_j)(z_2-e^\epsilon w_j)}~,}
and
\eqn\rkdef{\eqalign{r_k={w_k\over(w_k-z_2)(e^\epsilon z_2-e^{-\epsilon} w_k)}
         &\prod_{i=1\atop i\ne k}^M
        {w_k^2 e^{-\epsilon}-w_i^2 e^\epsilon\over
         (w_ke^{-\epsilon}-w_i e^\epsilon)(w_k-w_i)}\cr
     \times\prod_{j=M+1}^{2m}&
      {z_2^2-w_j^2\over (z_2-e^{-\epsilon}w_j)(z_2-e^\epsilon w_j)}~,}}
for $1\le k\le M$.

$r_k$ and $r_z$ both appear to have poles on the $z_2$ contour when
$z_2=w_k$ or $w_i$, respectively.  However, we started with a completely
well-defined, convergent integral for any value of $z_2$ and $w_i$, so we
know that the result, $r_2+\sum_k r_k$, must be finite.  Consequently, all
the poles on the $z_2$ contour must cancel each other when we add up all the
residues.  (We have checked this explicitly for $m\le 2$.)  Therefore, we
can integrate each residue separately and just ignore the poles on the
contour.  Then, for $\oint r_z {dz_2\over 2\pi i}$, we must evaluate
residues when $z_2=0$ and $z_2 = e^{-\epsilon}w_l$ for $M+1\le l\le 2m$.
The residue when $z_2=0$ is
\eqn\rzzdef{r_{zz}={1\over e^\epsilon-e^{-\epsilon}}~.}
The residue when $z_2=e^{-\epsilon}w_l$ is
\eqn\rzldef{\eqalign{r_{zl}={1\over e^\epsilon-e^{-\epsilon}}
           &\prod_{i=1}^M{w_l^2e^{-2\epsilon}-w_i^2 e^{2\epsilon}\over
            (w_le^{-\epsilon/2}-w_i e^{\epsilon/2})
            (w_le^{-3\epsilon/2}-w_i e^{3\epsilon/2})}\cr
 \times\prod_{j=M+1\atop j\ne l}^{2m}&{w_l^2 e^{-\epsilon}-w_j^2e^\epsilon
       \over(w_l-w_j)(w_le^{-\epsilon}-w_je^\epsilon)}~.}}

The integral $\oint r_k {dz_2\over 2\pi i}$ has poles inside the contour
when $z_2=e^{-2\epsilon}w_k$ and $z_2=e^{-\epsilon}w_l$ for $M+1\le l\le 2m$.
The residue when $z_2=e^{-2\epsilon}w_k$ is
\eqn\rkkdef{\eqalign{r_{kk}= {1\over e^\epsilon-e^{-\epsilon}}
          &\prod_{i=1\atop i\ne k}^M
        {w_k^2 e^{-\epsilon}-w_i^2 e^\epsilon\over
         (w_ke^{-\epsilon}-w_i e^\epsilon)(w_k-w_i)}\cr
    \times\prod_{j=M+1}^{2m}&{w_k^2e^{-2\epsilon}-w_j^2 e^{2\epsilon}\over
            (w_ke^{-\epsilon/2}-w_j e^{\epsilon/2})
            (w_ke^{-3\epsilon/2}-w_j e^{3\epsilon/2})}~;}}
and, lastly, the residue when $z_2=e^{-\epsilon}w_l$ for $M+1\le l\le 2m$ is
\eqn\rkldef{\eqalign{r_{kl}={w_kw_l\over(w_ke^{\epsilon/2}-w_le^{-\epsilon/2})
                (w_le^{\epsilon/2}-w_ke^{-\epsilon/2})}&
          \prod_{i=1\atop i\ne k}^M
        {w_k^2 e^{-\epsilon}-w_i^2 e^\epsilon\over
         (w_ke^{-\epsilon}-w_i e^\epsilon)(w_k-w_i)}\cr
       \times \prod_{j=M+1\atop j\ne l}^{2m}&
           {w_l^2e^{-\epsilon}-w_j^2e^\epsilon
        \over (w_l-w_j)(w_le^{-\epsilon}-w_je^\epsilon)}~.}}

The total $2m$-point function is then given by a sum over all the residues,
symmetrized in the $w_k$'s:
\eqn\Atwosum{A_2=(-1)^m\sum_{M=0}^{2m}\sum_{\sigma_M}(-1)^M
        \left[r_{zz}+\sum_{l\notin\sigma_M}r_{zl} +\sum_{k\in\sigma_M}r_{kk}
        +\sum_{k\in\sigma_M}\sum_{l\notin\sigma_M}r_{kl}\right]~.}

To evaluate $A_2$ as $\epsilon\to0$, we Taylor expand $r_{zz}$, $r_{zl}$,
$r_{kk}$ and $r_{kl}$.  After some algebra, we find
\eqn\rzzexp{r_{zz}={1\over e^\epsilon-e^{-\epsilon}}
           ={1\over 2\epsilon}+O(\epsilon)~;}
\eqn\rzlexp{r_{zl}={1\over 2\epsilon}f(w_l)
                 +2\sum_{p\in\sigma_M}h(w_l,w_p)
                 +\sum_{p\notin\sigma_M\atop p\ne l}h(w_l,w_p)+O(\epsilon)~;}
\eqn\rkkexp{r_{kk}={1\over 2\epsilon}f(w_k)
                 +\sum_{p\in\sigma_M\atop p\ne k}h(w_k,w_p)
                 +2\sum_{p\notin\sigma_M}h(w_k,w_p)+O(\epsilon)~;}
and
\eqn\rklexp{r_{kl}=-H_{\sigma_M}(k,l)~,}
where we have defined
\eqn\fdef{f(w_l)=\prod_{i=1\atop i\ne l}^{2m}{w_l+w_i\over w_l-w_i}~;}
\eqn\hdef{h(w_k,w_p)={w_k w_p\over (w_k-w_p)^2}
             \prod_{i=1\atop i\ne k, i\ne p}^{2m}{w_k+w_i\over w_k-w_i}~;}
and
\eqn\Hsmdef{H_{\sigma_M}(k,l)={w_k w_l\over (w_k-w_l)^2}
             \prod_{i\in\sigma_M\atop i\ne k}{w_k+w_i\over w_k-w_i}
             \prod_{j\notin\sigma_M\atop j\ne l}{w_l+w_j\over w_l-w_j}~.}
We will also define the functions
\eqn\Fdef{F=\sum_{j=1}^{2m}f(w_j)~;}
\eqn\gdef{g(w_k)=\sum_{p=1\atop p\ne k}^{2m}h(w_k,w_p)~;}
and
\eqn\Gdef{G=\sum_{k=1}^{2m}g(w_k)~.}
Then we can write $A_2$ as the sum of the following three terms.  The first
comes from $r_{zz}$ and is independent of all $w_j$'s and $M$:
\eqn\Azzdef{A_{zz}=(-1)^m\sum_{M=0}^{2m}\sum_{\sigma_M}(-1)^M r_{zz}~.}
The second term comes from $r_{zl}$ and $r_{kk}$ and depends only on $F$.
\eqn\AFdef{A_F=(-1)^m\sum_{M=0}^{2m}\sum_{\sigma_M}(-1)^M{1\over2\epsilon}F~.}
The last term depends on $h$, $g$, and $H$.  It is
\eqn\AHdef{\eqalign{A_H=(-1)^m\sum_{M=0}^{2m}&\sum_{\sigma_M}(-1)^M
           \biggl[\sum_{l\notin\sigma_M}\sum_{p\in\sigma_M} h(w_l,w_p)
            +\sum_{k\in\sigma_M}\sum_{p\notin\sigma_M} h(w_k,w_p)\cr
           &+\sum_{k=1}^{2m}g(w_k)
            -\sum_{k\in\sigma_M}\sum_{l\notin \sigma_M} H(k,l)\biggr]~.}}

First, we will evaluate $A_F$ by demonstrating that $F=0$.  $F$ is given by
\eqn\Ffull{F=\sum_{k=1}^{2m}\prod_{i=1\atop i\ne k}^{2m}
                        {w_k+w_i\over w_k-w_i}~.}
We can put all the terms in $F$ over a common denominator to obtain
\eqn\Fcomden{F=\prod_{i>j}{1\over w_i-w_j}
            \sum_k(-1)^k\left[\prod_{i=1\atop i\ne k}^{2m}(w_k+w_i)\right]
            \left[\prod_{i>j\atop i,j\ne k}(w_i-w_j)\right]~.}
Now we can use the fact that $\prod_{i>j=1;i,j\ne k}^{2m}(w_i-w_j)$ is the
discriminant, which equals the Van der Monde determinant. Consequently,
we can write it as
\eqn\vandermonded{\prod_{i>j\atop i,j\ne k}(w_i-w_j)= \det M_k~,}
where $M_k$ is a $(2m-1)$ by $(2m-1)$ matrix with
\eqn\Mkdef{(M_k)_{ij}=\cases{(w_i)^{(j-1)}  &for $i<k$~;\cr
                             (w_{i+1})^{(j-1)} &for $i\ge k$~.\cr}}
Then we can write $F$ as
\eqn\calc{F=\prod_{i>j}{1\over w_i-w_j}
      \left\{\sum_k(-1)^k\left[\prod_{i\ne k}(w_k+w_i)\right]\det M_k\right\}.}
The expression in curly brackets looks like the determinant of the matrix $A$
when it is calculated by
expanding the last row, where $A$ is given by
\eqn\matrixAdef{A=\pmatrix{1&1&\ldots&1\cr
                          w_1&w_2&\ldots&w_{2m}\cr
                          \vdots&\vdots&\ddots&\vdots\cr
                          w_1^{2m-2}&w_2^{2m-2}&\ldots&w_{2m}^{2m-2}\cr
                          \prod_{i\ne1}(w_1+w_i)&\prod_{i\ne2}(w_2+w_i)&
                          \ldots&\prod_{i\ne2m}(w_{2m}+w_i)\cr}.}
We can write the $k$th entry of the last row as
\eqn\tmrow{\prod_{i\ne1}(w_k+w_i)=w_k^{2m-2}\left[\sum_{i=1}^{2m}w_i\right]
             +w_k^{2m-4}\left[\sum_{i<j<l}w_iw_jw_l\right]+~\cdots~
             +1\left[\sum_{i=1}^{2m}\prod_{j\ne i}w_j\right].}
 From this equation we can see that the last row of $A$ is equal to a linear
combination of the first $2m-1$ rows of $A$, so $\det A=0$.  Therefore, since
$F\propto\det A$, $F$ is zero.  Then we can conclude that the contribution,
$A_F$, to the $2m$-point function is also zero.

Next, we will evaluate $A_{zz}$ by performing the sum
\eqn\Sdef{S=\sum_{M=0}^{2m}\sum_{\sigma_M}(-1)^M~.}
The sum over $\sigma_M$ is a sum over all ways to choose $M$ objects from a
set of $2m$ objects, so $S$ is
\eqn\Scalc{S=\sum_{M=0}^{2m}{2m\choose M}(-1)^M=(1-1)^{2m}=0~.}
Then
\eqn\Azzcalc{A_{zz}=-(-1)^m r_{zz}S=0~.}

Lastly, we must evaluate $A_H$. We will begin by proving the identity
\eqn\Hhid{\sum_{k\in\pi}H_\pi(k,l)=\sum_{k\in\pi}h(w_l,w_k)~,}
where $\pi$ is a subset of $\{1,\dots,2m\}$ containing $M$ elements, and
$l\notin\pi$.  Using the definitions of $H_\pi$ and $h$, we can write
\eqn\Hmess{\sum_{k\in\pi}H_\pi(k,l)=
            \left[w_l\prod_{j\notin\pi\atop j\ne l}{w_l+w_j\over w_l-w_j}
            \prod_{i\in\pi}{1\over(w_i-w_l)^2}\right]
            \sum_{k\in\pi}w_k
            \prod_{i\in\pi\atop i\ne k}{w_k+w_i\over w_k-w_i}(w_i-w_l)^2~;}
and
\eqn\hmess{\sum_{k\in\pi}h(w_l,w_k)=
            \left[w_l\prod_{j\notin\pi\atop j\ne l}{w_l+w_j\over w_l-w_j}
            \prod_{i\in\pi}{1\over(w_i-w_l)^2}\right]
            \sum_{k\in\pi}w_k
            \prod_{i\in\pi\atop i\ne k}(w_l+w_i)(w_l-w_i)~.}
 From these two equations, we can conclude that
$$\sum_{k\in\pi}H_\pi(k,l)=\sum_{k\in\pi}h(w_l,w_k)\qquad\hbox{if and only
if}$$
\eqn\redid{\sum_{k\in\pi}w_k
            \prod_{i\in\pi\atop i\ne k}{w_k+w_i\over w_k-w_i}(w_i-w_l)^2
           =\sum_{k\in\pi}w_k
            \prod_{i\in\pi\atop i\ne k}(w_l^2-w_i^2)~.}
Subtracting the right-hand side from the left-hand side of this equation and
simplifying, we find that the condition becomes
\eqn\subid{0=\left[2^{M-1}\prod_{i\in\pi}w_i(w_l-w_i)\right]
           \sum_{k\in\pi}(w_k-w_l)^{M-2}
           \prod_{i\in\pi\atop i\ne k}{1\over w_k-w_i}~.}
Because the expression in square brackets does not equal zero for arbitrary
values of $w$, the above expression is true if and only if
\eqn\Bdef{0=\sum_{k\in\pi}(w_k-w_l)^{M-2}
       \prod_{i\in\pi\atop i\ne k}{1\over w_k-w_i}=B~.}
To make the argument simpler, we will take $\pi=\{1,2,\dots,M\}$.  Then we can
write the expression on the right-hand side of the above equation as
\eqn\Bcalc{\eqalign{B=&\left(\prod_{i,j\in\pi\atop i>j}{1\over w_i-w_j}\right)
            \sum_{k\in\pi}(-1)^{M+k}(w_k-w_l)^{M-2}
            \prod_{i>j\in\pi\atop i,j\ne k}(w_i-w_j)\cr
           =&\prod_{i,j\in\pi \atop i>j}{1\over w_i-w_j}
            \sum_{k\in\pi}(-1)^{M+k}(w_k-w_l)^{M-2}
            \det M_k~,}}
where $M_k$ is an $(M-1)$ by $(M-1)$ matrix defined as in equation \Mkdef.  The
sum in the expression for
$B$ is just the row expansion of the following determinant:
\eqn\zerodet{\det\pmatrix{1&1&\ldots&1\cr
                  w_1&w_2&\ldots&w_{M}\cr
                  \vdots&\vdots&\ddots&\vdots\cr
                  w_1^{M-2}&w_2^{M-2}&\ldots&w_{M}^{M-2}\cr
                  (w_1-w_l)^{M-2}&(w_2-w_l)^{M-2}&\ldots&(w_M-w_l)^{M-2}\cr}
                  =0~.}
The last row in the matrix is a linear combination of the other rows,
so $B=0$.  As a result, we have shown that the identity in equation \Hhid\
is true.

Using this identity, we can write $A_H$ as
\eqn\AHGcalc{A_H=(-1)^m\sum_{M=0}^{2m}\sum_{\sigma_M}(-1)^M
\left[G+\sum_{k\in\sigma_M}\sum_{p\notin\sigma_M}h(w_k,w_p)\right].}
Because $G$ is independent of $M$ and $\sigma_M$, we can perform the sum
over $G$ using equations \Sdef\ and \Scalc\ to obtain
\eqn\AHhcalc{A_H=(-1)^m\sum_{M=0}^{2m}\sum_{\sigma_M}(-1)^M
            \sum_{k\in\sigma_M}\sum_{p\notin\sigma_M}h(w_k,w_p)~.}
When we sum over all subsets, $\sigma_M$, the pair $(w_k,w_p)$ will take on
all possible values, and each particular $(w_k,w_p)$ will occur
${2m-2\choose M-1}$ times.  Then
\eqn\finAcal{\eqalign{A_H=&(-1)^m\sum_{M=1}^{2m-1}{2m-2\choose M-1}(-1)^M
                 \sum_{k,p=1\atop k\ne p}^{2m}h(w_k,w_p)\cr
                =&(-1)^m(1-1)^{2m-2}G~,}}
so $A_H=0$ for $2m > 2$.  Therefore, the $2m$-point functions for $2m>2$ are
all zero as $\epsilon\to 0$ at order $V_0^2$.

The calculation for the $O(V_0^4)$ contribution to the four-point function is
similar but longer.  There are two key differences. The first is that we
must be careful to subtract out the disconnected diagrams. The second is
that we must also make use of the identity
$\sum_\sigma\prod_i 1/(w_{\sigma(i)}-w_{\sigma(i+1)})=0$, where the sum is over
all permutations, $\sigma$, of $2m$ elements, with $\sigma(2m+1)=\sigma(1)$.

In all the $m$-point function calculations we have done, the only function
that we have obtained that is SU(1,1) invariant is
$\prod_i w_i/(w_i-w_{i+1})$.  For example, $F$ and $G$ are not $SU(1,1)$
invariant.  Additionally, the $m$-point functions of $\dot X(t_i)$ must be
symmetric under interchange of the $t_i$ (at least when $\alpha=1$). We
find that when we symmetrize the SU(1,1) invariant function, we get $0$.
This suggests that $\langle\dot X(t_1)\dots\dot X(t_{2m})\rangle=0$ to
all orders in $V_0$, but does not guarantee it.

\appendix{B}{Ward Identity Check}

In Section 6, we noted that there is a complete set of ``broken
reparametrization invariance'' Ward identities which must be satisfied by
the free energy and correlation functions in the theory \CTrivi.  These
can be derived from the condition that the string theory boundary
state to which our theory corresponds must have reparametrization
invariance.  This condition is implemented by requiring \vira\ to
be satisfied, and in this appendix, for zero magnetic field, we check
the simplest form of this identity in order to show that our
renormalization scheme is correct.  It turns out that we need to know
the charged sector contributions to the free energy and to the two-point
function.  Both can be read off from the results in Section 6.

A special case of \vira\ is
\eqn\ward{(L_1-\tilde{L}_{-1})|B\rangle =0~,}
and it is shown in \CTrivi\ that this is equivalent to the following
condition on the connected diagram generating functional $W$:
\eqn\spward{\sum_{m=-\infty\atop m\ne 0}^\infty (m+1)\alpha_{-m}
            {\del W\over\del\alpha_{-m-1}} =
            {il\over{2}}\left[{\del^2 W\over\del X_0\del\alpha_{-1}}
            -{\del W\over\del\alpha_{-1}}
            {\del W\over\del X_0}\right]~,}
where the $\alpha_m$ are the closed string oscillators (see \SLSdef), and
$l=\sqrt{2\apm}$.
If we differentiate with respect to $\alpha_p$ and then set each
$\alpha_m = 0$ we get (using the fact that the one-point function
${\del W\over\del\alpha_{m}}=0$ on any solution)
\eqn\dward{\left.{\del\over\del X_0}
           {\del^2 W\over\del\alpha_p\del\alpha_{-1}}
           -{\del^2 W\over\del\alpha_p\del\alpha_{-1}}
           {\del W\over\del X_0}\right|_{\alpha_{m}=0} = 0~.}
Now ${\del^2 W\over\del\alpha_1\del\alpha_{-1}}$ is the first Fourier
mode of $\langle\dot{X}(\tau)\dot{X}(0)\rangle$, so \dward\ with $p=1$
is equivalent to (we take $T=2\pi$)
\eqn\ftward{{\del\over\del X_0}\int_0^{2\pi}{d\tau\over2\pi}
            e^{i\tau}\lb\dot{X}(\tau)\dot{X}(0)\rb
            = \left(\int_0^{2\pi}{d\tau\over2\pi}e^{i\tau}
            \lb\dot{X}(\tau)\dot{X}(0)\rb\right)
            {\del W\over\del X_0}~.}
This is the equation we must verify.
We now substitute the calculated forms for $W$ (equal to $T F$ when the
sources are set to zero) and
$\langle\dot{X}(\tau)\dot{X}(0)\rangle$ from equations \free, \tpi\ and
\tpii\ to check whether \ftward\ is satisfied.
The $m$th Fourier mode of the two-point function for $m>0$ is
\eqn\fttpt{\eqalign{\int_0^{2\pi}{d\tau\over2\pi}e^{im\tau}\lb
           \dot{X}(\tau)\dot{X}(0)\rb = & ~m - V_r\cos X_0\cr
           &+{V_r^2\over4}\left[2\cos 2X_0~\delta_{m,1}
		+4\sin^2 X_0-m\right]+~\cdots}}
and for $m=1$ the result is
\eqn\ftmone{\int_0^{2\pi}{d\tau\over2\pi}e^{i\tau}\lb
            \dot{X}(\tau)\dot{X}(0)\rb =
            1 - V_r\cos X_0 + \quart V_r^2+~\cdots~.}
The generating functional is given by
\eqn\wform{W=-V_r\cos X_0 - \half V_r^2\left({1\over4\epsilon}
					-\sin^2 X_0\right)+O(V_r^3)~,}
and it is easy to see that \ftward\ is satisfied to $O(V_r^2)$.  Note
that the contact terms played a crucial role in this calculation,
confirming that they are an essential part of the physical content of
the critical theory.  This Ward identity represents a very non-trivial
check that our renormalization scheme is consistent.

\listrefs
\bye